\documentclass[12pt]{article}
\pdfoutput=1
\usepackage{amsmath,amssymb,amsthm,bm,graphicx,float,array,multirow,multicol,rotfloat,caption,subcaption,hyperref,cleveref,enumerate,geometry,mathdots,adjustbox,booktabs,parskip,mathtools,tikz,tikz-cd,pdflscape,csquotes,lscape,rotating,empheq,mathrsfs}
\usepackage[para]{threeparttable}
\usepackage[all]{xy}
\usepackage[normalem]{ulem}
\usepackage[numbers,sort&compress]{natbib}
\usepackage{comment}

\usetikzlibrary{positioning}
\numberwithin{equation}{section}
\numberwithin{figure}{section}
\allowdisplaybreaks
\makeatletter

\usetikzlibrary{arrows,shapes.misc}
\tikzset{gauge/.style={rounded rectangle, draw=black!100,dashed, thick, minimum size=5mm},d2/.style={rounded rectangle, draw=white!100, thick, minimum size=5mm},flavor/.style={rectangle, draw=black!100, thick, minimum size=5mm},gaugeN1/.style={rounded rectangle, draw=black!100, thick, minimum size=5mm}}

\theoremstyle{plain}
\newtheorem*{thm*}{Theorem}

\theoremstyle{definition}

\newtheorem*{defn*}{Definition}

\makeatother

\newcommand{\trace}{\textrm{Tr}}

\newcommand{\CN}{\mathcal{N}}

\begin{document}

\begin{titlepage}
\vspace*{-3cm} 
\begin{flushright}
{\tt CALT-TH-2021-040}\\
{\tt DESY-21-190}\\
\end{flushright}
\begin{center}
\vspace{2cm}
{\LARGE\bfseries 
Infinitely many 4d $\mathcal{N}=1$ SCFTs with $a = c$\\}
\vspace{1.2cm}

{\large
Monica Jinwoo Kang$^{1,2}$, Craig Lawrie$^{3}$, Ki-Hong Lee$^2$, and Jaewon Song$^2$\\}
\vspace{.7cm}
{ $^1$ Walter Burke Institute for Theoretical Physics, California Institute of Technology}\par
{Pasadena, CA 91125, U.S.A.}\par
\vspace{.2cm}
{ $^2$ Department of Physics, Korea Advanced Institute of Science and Technology}\par
{Daejeon 34141, Republic of Korea}\par
\vspace{.2cm}
{ $^3$ Deutsches Elektronen-Synchrotron DESY}\par
{Notkestr.~85, 22607 Hamburg, Germany}\par
\vspace{.2cm}

\vspace{.3cm}

\scalebox{.8}{\tt monica@caltech.edu, craig.lawrie1729@gmail.com, khlee11812@gmail.com, jaewon.song@kaist.ac.kr}\par
\vspace{1.2cm}
\textbf{Abstract}
\end{center}

\noindent We study a rich set of four-dimensional $\mathcal{N}=1$ superconformal field theories (SCFTs) with both central charges identical: $a = c$. We construct them via the diagonal $\mathcal{N}=1$ gauging of the flavor symmetry $G$ of a collection of $\mathcal{N}=2$ Argyres--Douglas theories of type $\mathcal{D}_p(G)$, with or without additional adjoint chiral multiplets. In this way, we construct infinitely-many theories that flow to interacting SCFTs with $a = c$ in the infrared. Finally, we briefly highlight the features of the SCFTs without $a = c$ that arise from generalizing this construction.

\vfill 
\end{titlepage}

\tableofcontents

\newpage


\section{Introduction}

When a conformal field theory in four-dimensions is put on a curved manifold, the conformal symmetry becomes anomalous and characterized by two quantities $a$ and $c$, commonly referred to as the central charges. The trace of the energy-momentum tensor is non-vanishing and is given by 
\begin{align}
     T_\mu{}^\mu = c W^2 - a E_4 \ , 
\end{align}
where $W^2$ is the square of the Weyl tensor and $E_4$ is the Euler density. It has been shown that the central charge $a$ is a monotonically decreasing function along renormalization group flow \cite{Komargodski:2011vj}
\begin{align}
    a_{\mathrm{IR}} < a_{\mathrm{UV}} \ , 
\end{align}
which justifies the intuitive notion of treating the central charge $a$ as a measure for counting the degrees of freedom in a field theory. On the other hand, the central charge $c$ is \emph{not} a monotonically decreasing function along the RG flow. In many examples (especially with higher degrees of supersymmetry), the $c$ function mostly decreases along the flow, but there are examples in which we get $c_{\mathrm{IR}} > c_{\mathrm{UV}}$.\footnote{A simple example in the SUSY setup where $c$ does not decrease along the RG flow can be constructed by considering $\mathcal{N}=1$ $SU(2)$ gauge theory with one adjoint and two fundamental chiral multiplets. See the flow from $\mathcal{T}_0$ theory to $H_0$ theory in \cite{Maruyoshi:2018nod}.}                      

Unitarity puts constraints on the ratio of the central charges $a/c$ to be \cite{Hofman:2008ar,Hofman:2016awc}
\begin{align}
    \frac{1}{3} \le \frac{a}{c} \le \frac{31}{18} \,, 
\end{align}
where the lower bound is saturated by the theory of a free scalar boson and the upper bound is saturated by the free vector. For supersymmetric theories, the bound becomes narrower: 
\begin{align}
    \mathcal{N}=1~\textrm{SCFTs}\ :\quad \frac{1}{2} \le \frac{a}{c} \le \frac{3}{2} \,, \\
    \mathcal{N}=2~\textrm{SCFTs}\ :\quad \frac{1}{2} \le \frac{a}{c} \le \frac{5}{4} \,. 
\end{align}
The lower bound is saturated by a free chiral multiplet or a free hypermultiplet and the upper bound is saturated by an $\mathcal{N}=1$ and $\mathcal{N}=2$ vector multiplet, respectively. With higher supersymmetry, it is known that the central charges are equal \cite{Aharony:2015oyb}: 
\begin{align}
    \mathcal{N}=3, 4~\textrm{SCFTs}\ :\quad a = c \,.
\end{align}

Apart from $\mathcal{N}=3, 4$ superconformal theories, the central charges $a$ and $c$ are not generally related to each other. However, for the theories that are holographically dual to AdS$_5$ gravity, the two central charges are equal (\textit{i.e.}~$a=c$) in the large $N$ limit \cite{Henningson:1998gx}. For a finite $N$, the difference of the central charges $(a-c)$ gives rise to the $R^{\mu\nu\rho\sigma}R_{\mu\nu\rho\sigma}$ correction in the effective gravity action \cite{Henningson:1998gx,Anselmi:1998zb}. This results in corrections to the celebrated entropy-viscosity ratio bound \cite{Kovtun:2004de} to give \cite{Kats:2007mq, Buchel:2008vz}
\begin{align}
    \frac{\eta}{s} \ge \frac{1}{4\pi} \left(1 - \frac{c-a}{c} + \cdots \right) \ . 
\end{align}
The difference between the central charges $(a-c)$ also controls universal behavior of certain quantities of SCFTs, such as the Cardy-like limit of the superconformal index \cite{DiPietro:2014bca} and the entanglement entropy \cite{Perlmutter:2015vma}. This difference $(a-c)$ also controls phenomena in CFT and holography such as the mixed current-gravitational anomaly \cite{Anselmi:1997am} and the size of the single-trace higher spin gap for large $N$ \cite{Camanho:2014apa}.

Then, a question to ponder would be whether one can have theories with $a=c$ even in finite $N$ without a high degree of supersymmetry and which, in turn, would comprise a rather special set of CFTs. In the context of 4d $\mathcal{N}=2$ SCFTs, families of such theories with $a=c$ have been found to exist in \cite{Kang:2021lic}. What we find in this paper is that 4d SCFTs with $a=c$ exist with the minimal amount of supersymmetry (namely, $\mathcal{N}=1$) as well and the set of such theories is quite broad. 

An interesting set of 4d $\mathcal{N}=2$ SCFTs is studied in \cite{Kang:2021lic} called $\widehat{\Gamma}(G)$, associated to a choice of an affine ADE Dynkin diagram $\widehat{\Gamma}$ and an ADE gauge group $G$. These theories are obtained via gauging all of the $G$ flavor symmetries of a collection of Argyres--Douglas and conformal matter theories. For a choice of $\Gamma = D_4, E_6, E_7, E_8$ and particular choices of $G$, there is a fascinating connection to $\mathcal{N}=4$ super Yang--Mills (SYM) theory, whereby their central charges are identical,
\begin{equation}
    a(\widehat{\Gamma}(G)) = c(\widehat{\Gamma}(G)) \,,
\end{equation}
and their Schur indices can be rewritten in terms of the Schur index of $\mathcal{N}=4$ SYM,
\begin{equation}
    I_{\widehat{\Gamma}(G)}(q) = I^{\mathcal{N}=4}_G(q^{\alpha_\Gamma}, q^{\frac{\alpha_\Gamma}{2}-1}) \,,
\end{equation}
where $\alpha_\Gamma$ is the maximal coroot of $\widehat{\Gamma}$ \cite{Kang:2021lic}. This relationship occurs when\footnote{In fact, this relationship between the Schur indices holds whenever the SCFT $\widehat{\Gamma}(G)$ has no flavor symmetry. Equation \eqref{eqn:gcd1} is a sufficient, but not necessary, condition for the absence of flavor.}
\begin{equation}\label{eqn:gcd1}
    \gcd(\alpha_\Gamma, h_G^\vee) = 1 \,.
\end{equation}

A natural question is then whether there exist $\mathcal{N}=1$ theories which have a similar connection to $\mathcal{N}=4$ SYM, as we just witnessed for $\mathcal{N}=2$ theories. 
We can easily generate a set of $\mathcal{N}=1$ SCFTs by deforming $\widehat{\Gamma}(G)$ theories via a mass term for the $\mathcal{N}=1$ adjoint chiral multiplet $\phi$ in the $\mathcal{N}=2$ vector multiplet of $G$:
\begin{align}
    W=\frac{1}{2} m \trace\phi^2 \,.
\end{align}
Upon renormalization group flow (below the mass scale set by $m$), the mass-deformed theory flows to an $\mathcal{N}=1$ superconformal theory with the central charges given by \cite{Tachikawa:2009tt}
\begin{align}\label{eqn:coolratio}
    a_{\mathcal{N}=1} = \frac{27}{32} a_{\mathcal{N}=2} \ , \quad c_{\mathcal{N}=1} = \frac{27}{32} c_{\mathcal{N}=2} \ , 
\end{align}
which means that the mass-deformed $\widehat{\Gamma}(G)$ theories also have equal central charges $a=c$. The mass term generates a marginal coupling of the form
\begin{align} \label{eq:mumu}
    W= \trace \left(\sum_i \mu_i\right)^2 \ , 
\end{align}
where we omit the coupling constant. The mother $\mathcal{N}=2$ SCFT has a marginal gauge coupling, and the mass deformed $\mathcal{N}=1$ theory also has an exactly marginal coupling given by this term \cite{Leigh:1995ep, Green:2010da}. This is a direct analog of the quartic coupling of $SU(N)$ SQCD with $2N$ flavors that appears in many contexts. 

\begin{figure}[H]
    \centering
    \vspace{5mm}
    \begin{tikzcd}[column sep=0.1in, row sep=0.3in,labels={font=\normalsize}]
    \mathcal{N}=2\ \widehat{\Gamma}(G)\text{ theory}\arrow[rightarrow]{rdd}[description,pos=0.5]{\trace\phi^2} \arrow[rightarrow]{rrd}[red,font=\LARGE,description,pos=0.5]{\bigtimes}\\
    & & \CN=1~W=0\text{ fixed point}\arrow[rightarrow]{ld}[description,pos=0.5]{\trace\mu_i\mu_j}\\
    & \text{Mass deformed } \widehat{\Gamma}(G)\text{  theory}&
    \end{tikzcd}
    \caption{RG flows triggered by the mass deformation of the $\CN=2$ theory vs gauging an $\CN=1$ vector multiplet. 
    The red cross is to emphasize that there is no direct renormalization group flow between the $\mathcal{N}=2$ gauging and the $\mathcal{N}=1$ gauging.}
    \label{fig:mdflow}
\end{figure}
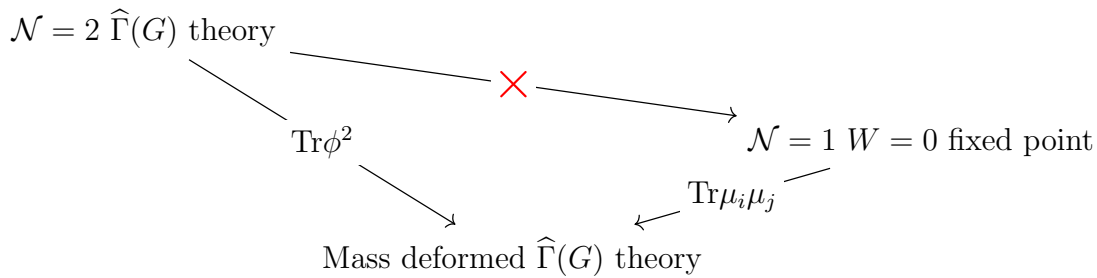

We also obtain an $\mathcal{N}=1$ theory with $a=c$ by simply replacing the $\mathcal{N}=2$ gauge multiplet in the $\widehat{\Gamma}(G)$ construction by an $\mathcal{N}=1$ gauge multiplet. We may expect that this theory is identical to the mass-deformation of the mother $\mathcal{N}=2$ theory with the famous $27/32$ ratio of central charges up on marginal deformation. However, we see that this is not the case! In fact, what we find in general is that there exists a $W=0$ fixed point (without any superpotential), at which the central charge $a_{\mathcal{N}=1}$ is bigger than $(27/32)\,a_{\mathcal{N}=2}$, and where some  operators of the form $\trace\mu_i \mu_j$ are \emph{relevant}. Upon deforming by these relevant operators, the theory flows to the mass-deformed theory with the central charge ratio given by $27/32$. This is reminiscent of the phenomenon that appears in $\mathcal{N}=1$ class $\mathcal{S}$ theory \cite{Benini:2009mz, Bah:2012dg, Bah:2013aha}, where the mass deformed theory corresponds to the case with equal degrees of normal bundles of more general compactifications. We depict such RG flows in Figure \ref{fig:mdflow}. 
Besides these theories obtained via simple (universal) relevant deformation of $\widehat{\Gamma}(G)$ theories, we find there exist a wider class of theories with $a=c$ that have no direct $\mathcal{N}=2$ origin. Such an $\mathcal{N}=1$ theory with $a=c$ can be constructed via gauging several $\mathcal{D}_{p_i}(G)$ theories with an $\mathcal{N}=1$ vector multiplet and possibly additional chiral multiplets in the adjoint representation of $G$. For some special choices, we can reproduce the mass deformed $\widehat{\Gamma}(G)$ theory. 

To verify that the 4d $\mathcal{N}=1$ SCFTs that we obtain in this manner truly are interacting SCFTs with $a = c$, it is necessary to check that they do not have a decoupled free sector, after flowing into the infrared. In this paper, we check that the Coulomb branch operators and products of the moment maps in the gauged theory satisfy the necessary unitarity conditions. A more refined test of unitary involves the computation of the superconformal index; in \cite{OPSPEC} we perform this computation and confirm that the $a = c$ theories constructed herein are indeed interacting SCFTs. The index also aids in the identification of the relevant operators of the theory; superpotential deformations via these operators may then trigger a flow to a new infrared SCFT. Intriguingly, we find that many of these deformations preserve the $a = c$ property, and we will explore the landscape of such deformations in \cite{LANDSCAPE}.

The structure of this paper is as follows. In Section \ref{sec:Neq1gluings}, we explain how and when the $\mathcal{N}=1$ gaugings of a collection of $\mathcal{D}_{p_i}(G)$ theories lead to infrared SCFTs with $a = c$. In Section \ref{sec:acgauging}, we show precisely under which conditions any such gauging will flow to a theory with identical central charges, and in Section \ref{sec:ps}, we determine which particular combinations of $p_i$ can be gauged in an asymptotically-free or conformal manner. In Sections \ref{sec:1DpG}, \ref{sec:2DpG}, \ref{sec:3DpG}, \ref{sec:4DpG}, and \ref{sec:5DpG}, we construct the infrared theories obtained from all the asymptotically free gaugings, with one to five $\mathcal{D}_p(G)$, respectively; and in Section \ref{sec:6DpG}, we analyze one instance of a conformal gaugings which involves six $\mathcal{D}_2(G)$.  We study if the constructed infrared theories are superconformal field theories and whether they satisfy the unitarity requirements to have $a = c$. In Section \ref{sec:withchirals}, we consider gauging $\mathcal{D}_{p_i}(G)$ together with additional matter charged under $G$. We explore gaugings with an additional adjoint chiral multiplet in Section \ref{sec:N2pot}, and we relate those to the mass-deformed $\widehat{\Gamma}(G)$ theories in Section \ref{sec:massdef}. In Section \ref{sec:lagrangian}, we consider a simple Lagrangian SCFT with $a = c$ which has two adjoint chirals and no $\mathcal{D}_p(G)$, and then in Section \ref{sec:2adjoints}, we enumerate all possible gaugings of $\mathcal{D}_{p}(G)$s with two adjoint chirals such that the infrared SCFT has $a = c$. Finally, in Section \ref{sec:beyondac}, we consider a generalization to gaugings of collections of $\mathcal{D}_p(G)$ theories and $(G, G)$ conformal matter theories. We summarize and suggest some future directions in Section \ref{sec:discussion}.

\section{\texorpdfstring{\boldmath{$\mathcal{N}=1$}}{N=1} gluing of \texorpdfstring{\boldmath{$\mathcal{D}_p(G)$}}{Dp(G)} theories}
\label{sec:Neq1gluings}

We are interested in constructing 4d $\mathcal{N}=1$ superconformal field theories starting from the non-Lagrangian 4d $\mathcal{N}=2$ SCFTs known as the $\mathcal{D}_p(G)$ Argyres--Douglas theories \cite{Cecotti:2012jx,Xie:2012hs,Cecotti:2013lda,Wang:2015mra}.\footnote{Some of the $\mathcal{D}_p(G)$ theories are actually $\mathcal{N}=2$ Lagrangian quivers, and sometimes they admit an $\mathcal{N}=1$ Lagrangian description \cite{Maruyoshi:2016tqk,Maruyoshi:2016aim,Agarwal:2016pjo,Agarwal:2017roi,Benvenuti:2017bpg}. We remind the reader that the $\mathcal{D}_p(SU(2))$ theory is identical to the $(A_1, D_{p})$ theory.} An interesting set of 4d $\mathcal{N}=2$ SCFTs, constructed out of $\mathcal{D}_p(G)$ theories and $(G\times G)$ conformal matter, called $\widehat{\Gamma}(G)$, are studied in \cite{Kang:2021lic}, where $\Gamma$ and $G$ are algebras of type ADE.\footnote{When $\Gamma = D_4, E_6, E_7, E_8$ these theories are known as the elliptic $G$-models of \cite{Cecotti:2011rv}; some aspects of some of those particular theories have also been explored in \cite{Buican:2016arp,Buican:2020moo,DelZotto:2015rca,Closset:2020afy,Cecotti:2013lda}.} Specifically, the authors considered a diagonal gauging of the $G$ flavor symmetry of a collection of $\mathcal{D}_{p_i}(G)$ and $(G\times G)$ conformal matter theories in such a way that, for each introduced gauge node, the introduced gauge coupling has a vanishing one-loop $\beta$-function. In this paper, we consider an analogous construction for 4d $\mathcal{N}=1$ SCFTs via the diagonal gauging of the flavor symmetry $G$ of a collection of $\mathcal{D}_{p_i}(G)$.

There are several subtleties that arise in the $\mathcal{N}=1$ case that are absent in the $\mathcal{N}=2$ gauging. In the simplest setup, we consider gauging by including in the theory a single $\mathcal{N}=1$ vector multiplet (in the adjoint representation of $G$) and coupling the vector multiplet and the $G$-flavor currents.  
Post-gauging, we are interested in obtaining a superconformal field theory after performing the renormalization group flow into the infrared.
In order to obtain a non-trivial theory in the IR, the gauge coupling for $G$ must be asymptotically free. We could also consider the case where the one-loop $\beta$-function for the gauge coupling is exactly zero; however, we then require the existence of exactly marginal operators for the theory to be an interacting SCFT. Such operators are not necessarily guaranteed to exist; however, we find that in some examples they do exist. 
In this paper, we will not study the conformal gaugings in detail, however, the properties of the infrared SCFTs, when they exist, can be determined via a straightforward application of the methods discussed herein. Furthermore, asymptotic freedom is not sufficient to guarantee the existence of an interacting SCFT in the infrared; we must check that there exists a non-anomalous superconformal R-symmetry. The superconformal R-symmetry in the infrared can be determined via $a$-maximization \cite{Intriligator:2003jj}, and involves mixing between the R-symmetry in the ultraviolet and (possibly emergent) Abelian flavor symmetries of the theory.

\subsection{\texorpdfstring{$a = c$}{a=c} for \texorpdfstring{$\mathcal{N}=1$}{N=1} gauging}
\label{sec:acgauging}

One of the fascinating features of the $\widehat{\Gamma}(G)$ theories, which are the 4d $\mathcal{N}=2$ SCFTs studied in \cite{Kang:2021lic}, is that a number of cases have their central charges being identical: $a = c$. In this section, we prove that many of the $\mathcal{N}=1$ gaugings share this feature. 

The $\mathcal{N}=1$ gauging breaks the $\mathcal{N}=2$ R-symmetry of the $\mathcal{D}_p(G)$ theory to the $U(1)$ R-symmetry and a flavor $U(1)$ symmetry, with generators 
\begin{align}
    R_0 =\frac{1}{3}R_{\mathcal{N}=2}+\frac{4}{3}I_3\ ,\quad
    \mathcal{F}=-R_{\mathcal{N}=2}+2I_3\ ,
\end{align}
where $R_{\mathcal{N}=2}$ and $I_3$ respectively denote the $U(1)_R$ charge and the $SU(2)_R$ Cartan of the $\mathcal{N}=2$ R-symmetry. Upon gauging, only the anomaly-free combinations are preserved. At the IR fixed point, the flavor $\mathcal{F}_i$ of each $\mathcal{D}_{p_i}(G)$ can mix with the R-symmetry before gauging 
\begin{align}
\begin{split}
    R=R_0+\sum_i \epsilon_i \mathcal{F}_i\ ,
\end{split}
\label{eq:IR_R}
\end{align}
and thus the $R$-charge gets modified in the infrared. Now we have to require that the superconformal R-symmetry at the fixed point satisfies the anomaly cancellation condition\footnote{When considering anomalies in 4d theories, each trace term of $GGG$, $FGG$, $FFG$, and $FFF$ respectively corresponds to the gauge anomaly, the ABJ anomaly of the flavor $F$, a mixed gauge-global anomaly that naturally cancels if $G$ is non-Abelian, and the 't Hooft anomaly. The gauge anomaly trivially vanishes since the theory has chiral symmetry. Here, we are considering the ABJ anomaly of the R-symmetry after gauging, which must vanish if it flows in the IR limit to an interacting SCFT. The mixed anomaly for an Abelian gauge theory can be non-vanishing and is responsible for a 2-group global symmetry \cite{Cordova:2018cvg}.}
\begin{align}
    0=\trace RGG=h^\vee_G+\sum_i\left(\left(\frac{1}{3}-\epsilon_i\right)\trace_iR_{\mathcal{N}=2}GG+\left(\frac{4}{3}+2\epsilon_i\right)\trace_iI_3GG\right),
\label{eq:RGG}
\end{align}
where $i$ runs through all the $p_i$ of $\mathcal{D}_{p_i}(G)$, which is equivalent to the vanishing condition of the Novikov--Shifman--Vainshtein--Zakharov (NSVZ) exact $\beta$-function for the gauge coupling.
The first $h^\vee_G$ term in equation \eqref{eq:RGG} is the contribution from the gauginos and for the next terms in equation \eqref{eq:RGG} we utilize the flavor central charge $k_G$ of the $\mathcal{D}_{p}(G)$ theories \cite{Cecotti:2013lda}:
\begin{align}
    \trace R_{\mathcal{N}=2}GG=-\frac{1}{2}k_G=-\frac{p-1}{p}h^\vee_G\ .
\end{align}
Then we are left with the relation involving the mixing coefficients $\epsilon_i$,
\begin{align}
    h^\vee_G+\sum_i\left(\frac{1}{3}-\epsilon_i\right)\left(-\frac{p_i-1}{p_i}h^\vee_G\right)=0 \,,
\end{align}
where the exact values of the $\epsilon_i$ are fixed by $a$-maximization \cite{Intriligator:2003jj}.
There are situations where there are more $U(1)$ flavor symmetries than the ones we considered above (namely the unbroken $U(1)$ symmetries inside the $\mathcal{N}=2$ R-symmetry).  
For example, when $G=SU(N)$ and $\mathcal{D}_{p}(G)$ has extra flavor symmetry beyond $G$, the $\mathcal{D}_p(G)$ theory is a Lagrangian quiver gauge theory and the additional flavor symmetries are the $U(1)$s rotating the bifundamental hypermultiplets. As these $U(1)$s are baryonic, they do not mix with the R-symmetry \cite{Intriligator:2003jj}.

A novel feature of the glued $\mathcal{N}=1$ theories is that they all have their two central charges to be equal, $a=c$, for particular combinations of $G$ and $p_i$. It follows from the evaluation of the difference of the central charges, and it does not rely on the knowledge of the exact superconformal R-symmetry,
\begin{align}\label{eq:trR}
    \begin{split}
    \trace R&=16(a-c)\\
    &=\textrm{dim}(G)+\sum_i\left(\left(\frac{1}{3}-\epsilon_i\right) \trace_i R_{\mathcal{N}=2}+\left(\frac{4}{3}+2\epsilon_i\right)\trace_iI_3\right)\\
    &=\textrm{dim}(G)+\sum_i\left(\frac{1}{3}-\epsilon_i\right)48(a_i-c_i)\ ,
    \end{split}
\end{align}
where $a_i$ and $c_i$ are the central charges of each $\mathcal{D}_{p_i}(G)$ theory. Here we used the relation \cite{Anselmi:1997am}
\begin{align} \label{eq:ac}
    a = \frac{3}{32}(3\trace R^3 - \trace R) \ , \quad c = \frac{1}{32}(9 \trace R^3 - 5 \trace R) \ .
\end{align}
The central charges of the $\mathcal{D}_{p_i}(G)$ theories with $p_i$ and $h^\vee_G$ coprime (which comprises a subset of the $\mathcal{D}_p(G)$ theories without any extra flavor symmetry beyond $G$) simplify to
\begin{align}
\begin{aligned}
& a_i =\frac{1}{48}\frac{(4p_i-1)(p_i-1)}{p_i}\textrm{dim}(G)\ ,\\
& c_i =\frac{1}{12}(p_i-1)\textrm{dim}(G)\ .
\end{aligned}
\end{align}
Thus, if we assume that each $p_i$ is coprime to $h_G^\vee$, the equation \eqref{eq:trR} further implies the difference in the two central charges to be zero:
\begin{align}\label{eq:cmaRGG}
    \begin{split}
        a-c&=\frac{1}{16}\left(\textrm{dim}(G)+\sum_i\left(\frac{1}{3}-\epsilon_i\right)\left(-\frac{p_i-1}{p_i}\textrm{dim}(G)\right)\right)\\
        &=\frac{\textrm{dim}(G)}{16h^\vee_G}\ \trace RGG=0\, .
    \end{split}
\end{align}
This means that for these theories we have the two central charges $a$ and $c$ to be equal and the $\trace R$ term, presented in equation \eqref{eq:trR}, vanishes. We note that this argument is unchanged under a superpotential deformation of the 4d $\mathcal{N}=1$ theory and thus that the $a = c$ property is preserved under such deformations. This is because the superpotential deformation only puts further constraints on the $\epsilon_i$ and does not alter the structure of equation \eqref{eq:cmaRGG}, as long as there is no accidental symmetry that mixes with $R$. Superpotential deformations of these gaugings will be explored in \cite{LANDSCAPE}. 

Let us make a remark that the condition that $p$ and $h^\vee_G$ are coprime is stronger than the condition for the $\mathcal{D}_p(G)$ theory to have no additional flavor symmetry beyond $G$. For example, $\mathcal{D}_p(SO(2N))$ theory with odd $p$ has no extra flavor symmetry, but it does not give identical central charges after gauging as it does not necessarily follow that $\textrm{gcd}(p,2N-2) = 1$.

\subsection{Constraints on \texorpdfstring{$p_i$}{pi}}
\label{sec:ps}

We demonstrated in Section \ref{sec:acgauging} that if we consider a quiver gauge theory obtained by the diagonal gauging of the flavor symmetry $G$ of the collection of Argyres--Douglas theories $\mathcal{D}_{p_i}(G)$ such that $\prod_i \gcd(p_i, h_G^\vee) = 1$, then the $\mathcal{N} = 1$ superconformal field theories that the quiver may flow to in the infrared have $a = c$. 

As we are interested in the gauged $\mathcal{N}=1$ theories when they flow to superconformal field theories in the infrared, this requires that the gauge coupling for $G$ is asymptotically free. This puts strong constraints on the possible $p_i$ for which we can consider the gauging of $\mathcal{D}_{p_i}(G)$. 

Before we dive into the asymptotic-free condition for the $\mathcal{N}=1$ gauging, we revisit the $\mathcal{N}=2$ gauging for the theories explored in \cite{Kang:2021lic}. For those theories, it was crucial to consider the conformal gauging, as opposed to the asymptotically free gauging. Constructing a 4d $\mathcal{N}=2$ SCFT via conformal gauging of $N$ $\mathcal{D}_{p_i}(G)$ requires that
\begin{equation}\label{eqn:cfl_gauging}
    \sum_{i=1}^N \frac{2(p_i - 1)}{p_i}h_G^\vee = 4h_G^\vee \quad \Rightarrow \quad \sum_{i=1}^N \frac{1}{p_i} = N - 2 \,,
\end{equation}
where we have used that the central charge of the flavor symmetry of $\mathcal{D}_p(G)$ is 
\begin{equation}\label{eqn:kg}
    k_G = \frac{2(p - 1)}{p}h_G^\vee \,.
\end{equation}
It is straightforward to see that there are only four solutions (for finite values of $p_i$) to the Diophantine equation \eqref{eqn:cfl_gauging}, as shown in \cite{Cecotti:2013lda}:
\begin{equation}
    (p_1, p_2, p_3, p_4) = (2, 2, 2, 2), \, (1, 3, 3, 3),\, (1, 2, 4, 4) ,\, (1, 2, 3, 6) \,,
\end{equation}
where $\mathcal{D}_1(G)$ is the empty theory.\footnote{For convenience, we expressed the solutions with $i=1,2,3,4$.} These four solutions are associated to the affine ADE Dynkin diagrams $\widehat{D}_4$, $\widehat{E}_6$, $\widehat{E}_7$, and $\widehat{E}_8$, respectively \cite{Kang:2021lic}. 

A similar analysis can be performed to determine which $\mathcal{N}=1$ gaugings of $\mathcal{D}_{p_i}(G)$ are asymptotically-free, and thus which have a chance to flow to an interacting SCFT in the IR. Recalling the expressions for the flavor central charges in equation \eqref{eqn:kg}, we find that the asymptotic-free condition in this case is
\begin{equation}\label{eq:asymp_free}
    \sum_{i=1}^N \frac{2(p_i - 1)}{p_i}h_G^\vee\, <\, 6h_G^\vee\, .
\end{equation}
Since $h_G^\vee$ is nonzero, this inequality \eqref{eq:asymp_free} can be rewritten as a bound for a polynomial in $p_i$, as in the following Diophantine-like form:
\begin{equation}\label{eqn:asymp_alt}
    \sum_{i=1}^N \frac{1}{p_i} > N - 3 \,.
\end{equation}
To find all solutions, we need to analyze all possible values of $p_i \geq 2$ satisfying equation \eqref{eqn:asymp_alt}. It is clear that the maximum value of $N$ for which there exists such solutions is $N = 5$, and thus we consider each solution of equation \eqref{eqn:asymp_alt} as an unordered five-tuple 
\begin{align}
\label{eqn:5tuple}
    (p_1, p_2, p_3, p_4, p_5) \,,
\end{align}
where some of the $p_i$ may be one.\footnote{The $\mathcal{D}_1(G)$ refers to the empty theory, which can be simply ignored.} All tuples of $p_i$ which satisfy the asymptotic freedom condition in equation \eqref{eqn:asymp_alt} are listed in Table \ref{tbl:asympfreep}. We write these $\mathcal{N}=1$ gaugings via the quivers
\vspace{-4mm}
\begin{align}
\begin{aligned}
    \begin{tikzpicture}
      \node[gaugeN1] (s0) {$G$};
      \node[d2] (c1) [left=0.6cm of s0] {$\mathcal{D}_{p_3}(G)$};
      \node[d2] (c2) [right=0.6cm of s0] {$\mathcal{D}_{p_5}(G)$};
      \node[d2] (c3) [above=0.4cm of s0] {$\mathcal{D}_{p_4}(G)$};
      \node[d2] (c4) [below left=0.4cm of s0] {$\mathcal{D}_{p_2}(G)$};
      \node[d2] (c5) [below right=0.4cm of s0] {$\mathcal{D}_{p_1}(G)$};
      \draw (s0.east) -- (c2.west);
      \draw (s0.west) -- (c1.east);
      \draw (s0.north) -- (c3.south);
      \draw (s0.225) -- (c4.45);
      \draw (s0.315) -- (c5.135);
    \end{tikzpicture}
\end{aligned} \,,
\end{align}
where a solid-bordered node represents an $\mathcal{N}=1$ vector multiplet and a solid line between a $\mathcal{D}_p(G)$ theory and an $\mathcal{N}=1$ vector node corresponds to the inclusion of a term 
\begin{equation}\label{eqn:supcup}
     \mathcal{L} \sim \int d^4 \theta \mathcal{J}^a \mathcal{V}^a + \cdots \,,
\end{equation}
in the Lagrangian with $\mathcal{J}^a$ the flavor supercurrent from the $\mathcal{D}_p(G)$ SCFT and $\mathcal{V}^a$ the vector superfield.

We may be further interested in determining the possible tuples of $p_i$ such that the inequality in equation \eqref{eqn:asymp_alt} is saturated. When this occurs the one-loop $\beta$-function of the gauge coupling for $G$ vanishes directly. This does not generally lead to an interacting conformal field theory unless there is an exactly marginal operator. When we consider saturating the inequality there is one solution where one has six $p_i \geq 2$, and all other solutions have either four or five $p_i \geq 2$. All solutions are listed in Table \ref{tbl:confgaugep}. It is notable that solutions with six copies of $\mathcal{D}_2(G)$ possess sixteen exactly marginal operators, formed by products of the Coulomb branch operators, and thus they are interacting SCFTs with $a = c$.

\begin{table}[H]
\begin{threeparttable}
\centering
\renewcommand{\arraystretch}{1.15}
    $\begin{array}{ccccc}
    \toprule
      p_1 & p_2 & p_3 & p_4 & p_5 \\\midrule
      1 & 1 & 1 & 1 & p_5 \\
      1 & 1 & 1 & p_4 & p_5 \\
      1 & 1 & p_3 & p_4 & p_5 \\
      1 & 2 & 2 & p_4 & p_5 \\
      1 & 2 & 3 & \leq 6 & p_5 \\
      1 & 2 & 3 & 7 & \leq 41 \\
      1 & 2 & 3 & 8 & \leq 23 \\
      1 & 2 & 3 & 9 & \leq 17 
      \\\bottomrule
    \end{array}$
\quad
    $\begin{array}{ccccc}
    \toprule
      p_1 & p_2 & p_3 & p_4 & p_5 \\\midrule
      1 & 2 & 3 & 10 & \leq 14 \\
      1 & 2 & 3 & 11 & \leq 13\\
      1 & 2 & 4 & 4 & p_5 \\
      1 & 2 & 4 & 5 & \leq 19 \\
      1 & 2 & 4 & 6 & \leq 11 \\
      1 & 2 & 4 & 7 & \leq 9 \\
      1 & 2 & 5 & 5 & \leq 9 \\
      1 & 2 & 5 & 6 & \leq 7
      \\\bottomrule
    \end{array}$
\quad
    $\begin{array}{ccccc}
    \toprule
      p_1 & p_2 & p_3 & p_4 & p_5 \\\midrule
      1 & 3 & 3 & 3 & p_4 \\
      1 & 3 & 3 & 4 & \leq 11 \\
      1 & 3 & 3 & 5 & \leq 7 \\
      1 & 3 & 4 & 4 & \leq 5 \\
      2 & 2 & 2 & 2 & p_5 \\
      2 & 2 & 2 & 3 & 3 \\
      2 & 2 & 2 & 3 & 4 \\
      2 & 2 & 2 & 3 & 5
    \\\bottomrule
    \end{array}$
\end{threeparttable}
\caption{All possible tuples of $p_i$ such that $\mathcal{N}=1$ gauging of the common flavor symmetry of the associated $\mathcal{D}_{p_i}(G)$ leads to an asymptotically free theory. An entry that is left as $p_i$ indicates that the theory will be asymptotically free for any positive integer $p_i$.}
\label{tbl:asympfreep}
\end{table}

\begin{table}[H]
\begin{threeparttable}
\centering
\renewcommand{\arraystretch}{1.15}
    $\begin{array}{cccccc}
    \toprule
      p_1 & p_2 & p_3 & p_4 & p_5 & p_6 \\\midrule
      1 & 1 & 2 & 3 & 7 & 42 \\
      1 & 1 & 2 & 3 & 8 & 24 \\
      1 & 1 & 2 & 3 & 9 & 18 \\
      1 & 1 & 2 & 3 & 10 & 15 \\
      1 & 1 & 2 & 3 & 12 & 12 \\
      1 & 1 & 2 & 4 & 5 & 20
      \\\bottomrule
    \end{array}$
\quad
    $\begin{array}{cccccc}
    \toprule
      p_1 & p_2 & p_3 & p_4 & p_5 & p_6 \\\midrule
      1 & 1 & 2 & 4 & 6 & 12 \\
      1 & 1 & 2 & 4 & 8 & 8 \\
      1 & 1 & 2 & 5 & 5 & 10 \\
      1 & 1 & 2 & 6 & 6 & 6 \\
      1 & 1 & 3 & 3 & 4 & 12 \\
      1 & 1 & 3 & 3 & 6 & 6 
      \\\bottomrule
    \end{array}$
\quad
    $\begin{array}{cccccc}
    \toprule
      p_1 & p_2 & p_3 & p_4 & p_5 & p_6 \\\midrule
      1 & 1 & 3 & 4 & 4 & 6 \\
      1 & 1 & 4 & 4 & 4 & 4 \\
      1 & 2 & 2 & 3 & 3 & 3 \\
      1 & 2 & 2 & 2 & 4 & 4 \\
      1 & 2 & 2 & 2 & 3 & 6 \\
      2 & 2 & 2 & 2 & 2 & 2
    \\\bottomrule
    \end{array}$
\end{threeparttable}
\caption{The collections of $p_i$ that saturate the inequality in equation \eqref{eqn:asymp_alt}. For these choices of $p_i$, one obtains a theory where the one-loop $\beta$-function of the gauge coupling vanishes.}
\label{tbl:confgaugep}
\end{table}

Asymptotic freedom of the gauge coupling is a necessary condition for the existence of an interacting superconformal field theory at the infrared fixed point. However, we know that not all theories satisfying the asymptotic freedom condition, given by equation \eqref{eq:asymp_free}, flow to an interacting SCFT. Henceforth, we have to further check that the theory is consistent. To this end, we perform $a$-maximization \cite{Intriligator:2003jj} to determine the superconformal R-symmetry at the end of the RG flow. 
We then need to verify that the R-charges of the operators of the gauged theory under the superconformal R-symmetry satisfy the unitarity bounds.

If an operator appears to be non-unitary after the $a$-maximization, then somewhere along the flow into the infrared this operator becomes free and is decoupled from the theory \cite{Kutasov:2003iy}. One useful way to take care of the decoupled operator $\mathcal{O}$ is to introduce a flipper field $M$ with a superpotential coupling \cite{Barnes:2004jj, Benvenuti:2017lle, Maruyoshi:2018nod}:
\begin{align}
    W=M \mathcal{O} \,. 
\end{align}
In this way, one obtains the superconformal R-symmetry associated to the interacting sector of the infrared SCFT. This decoupling will violate the assumption made in Section \ref{sec:acgauging}, that we do not have any accidental $U(1)$ symmetries that can mix with the superconformal R-symmetry. For this case, the infrared SCFT will not have equal central charges.

We compute the superconformal R-symmetry via the procedure of $a$-maximization. 
Each $\mathcal{D}_{p_i}(G)$ theory has a moment map operator $\mu_i$ and a particular Coulomb branch operator with the smallest conformal dimension $u_i^0$. The operator $u_i^0$ is the bottom component of an $\CN=2$ chiral multiplet, which splits into two $\CN=1$ multiplets with bottom components $u_i^0$ and $Q^2 u_i^0$, where the $Q$ is an $\mathcal{N}=2$ supercharge with $(R_{\mathcal{N}=2},I_3)=(-1,1/2)$. The R-charges of these operators under the newfound infrared superconformal R-symmetry are
\begin{equation}
    R(\mu_i)=\frac{4}{3}+2\epsilon_i \,, \quad
    R(u^0_i)=\left(\frac{1}{3}-\epsilon_i\right)\frac{2(p_i+1)}{p_i} \,,\quad R\left(Q^2u_i^0\right)=\frac{4p_i+2}{3p_i}+\frac{2(p_i-1)}{p_i}\epsilon_i .
\end{equation}
Since we require $\trace \mu_i \mu_{j \neq i}$, $u_i^0$, and $Q^2u_i^0$ to be unitary operators of the SCFT at the IR fixed point, we get lower bounds on these R-charges as 
\begin{align}\label{eqn:unitarity}
    R(\mu_i)+R(\mu_{j\neq i}) > \frac{2}{3}\,,\quad R(u_i^0) > \frac{2}{3} \,,\quad R(Q^2u_i^0) > \frac{2}{3}\,. 
\end{align}
When these inequalities are saturated the theory is still unitary, but the associated operators become free and decouple from the theory. When this occurs the interacting sector of the resulting SCFT does not have $a = c$ because decoupled free chirals do not have identical central charges.
The second and the third conditions are obvious from the unitarity bound for the scalar ($\Delta \ge 1$), and the first term stems from the fact that the gauge-invariant operators of lowest dimension arise from $\trace \mu_i \mu_j$. In fact, $\trace \mu_i^k$ is zero in the chiral ring for all the $\mathcal{D}_p(G)$ theories with $(p, h^\vee)=1$. This can be understood from the fact that the Higgs branch for the $\mathcal{D}_p(G)$ theory is given by a nilpotent orbit of $G$ \cite{Arakawa2015, Song:2017oew}. Any nilpotent element $X$ in $G$ should have 
\begin{align} \label{eq:trmuk0}
    \trace X^k = 0\, .
\end{align}
Notice that $\langle \mu \rangle$ parametrize the Higgs branch. However, the operators of the form $\trace \mu_i \mu_j$ with $i \neq j$ are always present. 
The unitarity constraints in equation \eqref{eqn:unitarity} provide the upper and the lower bounds on the mixing parameters $\epsilon_i$ as 
\begin{align}
    -\frac{p_i+1}{3(p_i-1)}<\epsilon_i< \frac{1}{3(p_i+1)}\,,\quad \epsilon_i+\epsilon_{j\neq i} > -1\,.
\label{eq:unit_cond}
\end{align}
We will explore how a ``conformal window'' of possible sets of $p_i$ can be determined for gaugings of between one and five $\mathcal{D}_p(G)$ from these bounds on the mixing parameters $\epsilon_i$. It turns out that for the most of theories we consider, every $\epsilon_i$ satisfies
\begin{align}
-\frac{1}{3}<\epsilon_i < 0\,,    
\end{align}
which is a sufficient condition for equation \eqref{eq:unit_cond} to be satisfied.

Upon diagonally gauging a number of $\mathcal{D}_p(G)$ theories and flowing into the infrared, we find in the end that an infinite number of 4d $\mathcal{N}=1$ SCFTs with identical central charges, $a = c$, can be obtained. From Table \ref{tbl:asympfreep}, we pick any set of $(p_1, p_2, p_3, p_4, p_5)$ with either at most two of the $p_i$ being $1$, or else pick $(1, 1, 1, p_4, p_5)$ subject to the constraint that
\begin{equation}
    p_4\geq 3 \quad \text{and}\quad p_5\geq 3\,.
\end{equation}

We then consider the asymptotically-free theory obtained by coupling the flavor currents for the $G$ flavor symmetry of $\mathcal{D}_{p_i}(G)$, for each $p_i$, to an $\mathcal{N}=1$ vector multiplet.
When the flavor symmetry $G$ is such that $\gcd(p_i, h_G^\vee) = 1$, for all of the $p_i$, then the quiver that we have formed by the diagonal $\mathcal{N}=1$ gauging flows to an SCFT in the infrared with $a = c$.

\subsection{Gauging one \texorpdfstring{$\mathcal{D}_p(G)$}{Dp(G)} theory: no SCFT}
\label{sec:1DpG}

The simplest scenario of gauging is when we gauge $G$ of a single $\mathcal{D}_p(G)$ theory. The resulting $\mathcal{N}=1$ quiver is
\begin{align}
\begin{aligned}
    \begin{tikzpicture}
      \node[gaugeN1] (s0) {$G$};
      \node[d2] (c2) [left=0.6cm of s0] {$\mathcal{D}_{p}(G)$};
      \draw (s0.west) -- (c2.east);
    \end{tikzpicture}
\end{aligned}\ .
\end{align}
There is a unique anomaly-free R-charge, which is the generator of the superconformal R-symmetry in the IR:
\begin{align}
    \begin{split}
        R=R_0-\left(\frac{2}{3}+\frac{1}{p-1}\right)\mathcal{F} \quad \Rightarrow \quad \epsilon = - \left(\frac{2}{3} + \frac{1}{p-1}\right) \,.
    \end{split}
\end{align}
It is clear that this $\epsilon$ does not satisfy the unitarity condition, as it is not within the bounds in equation \eqref{eq:unit_cond}. The R-charges of the putative operators become
\begin{align}
    R(\mu) = -\frac{2}{p-1} \ , \quad R(u^0) = 2+\frac{4}{p-1} \ , \quad R(Q^2 u_0) = 0 \ .    
\end{align}
Hence, this theory does not flow to an interacting SCFT in the IR. When we have a negative R-charge operator, it is possible to generate a dynamic superpotential as in the celebrated case of Affleck--Dine--Seiberg (ADS) \cite{Affleck:1983mk}, where it was shown in particular for  $SU(N_c)$ SQCD with $N_f<N_c$ flavors. Even though the R-charge for $\mu$ is negative, all the gauge-invariant operators of the form $\trace \mu^k$ are not in the chiral ring; therefore we do not expect such a runaway dynamical superpotential is generated. Instead, we find a zero R-charge operator, $Q^2 u^0$, which is often responsible for quantum deformation of the vacuum moduli space as in the case of $SU(N_c)$ SQCD with $N_f=N_c$ flavors \cite{Seiberg:1994bz}. See also \cite{Maruyoshi:2013hja} for a similar phenomenon involving the non-Lagrangian $T_N$ theory.

\subsection{Gauging two \texorpdfstring{$\mathcal{D}_p(G)$}{Dp(G)} theories}
\label{sec:2DpG}

In contrast to considering a single $\mathcal{D}_p(G)$ theory, we can also form theories by gluing more than one $\mathcal{D}_p(G)$. We first consider gauging two $\mathcal{D}_p(G)$ theories together, which leads to a quiver of the following form:
\begin{align}
\begin{aligned}
    \begin{tikzpicture}
      \node[gaugeN1] (s0) {$G$};
      \node[d2] (c1) [left=0.6cm of s0] {$\mathcal{D}_{p_1}(G)$};
      \node[d2] (c2) [right=0.6cm of s0] {$\mathcal{D}_{p_2}(G)$};
      \draw (s0.east) -- (c2.west);
      \draw (s0.west) -- (c1.east);
    \end{tikzpicture} 
\end{aligned}\ .
\end{align}
As we determined in Section \ref{sec:ps}, the resulting theory is asymptotically free for arbitrary $(p_1, p_2)$, and we will assume without loss of generality that $p_1 \leq p_2$. Applying $a$-maximization, we discover that the superconformal R-symmetry is
\begin{equation}
    R = R_0 + \epsilon_1 \mathcal{F}_1 + \epsilon_2 \mathcal{F}_2 \,,
\end{equation}
where the two mixing parameters $\epsilon_1$ and $\epsilon_2$ are fixed by the values of $p_i$:
\begin{subequations}
\begin{align}
        \epsilon_1&=\frac{-p_1^2+p_2+p_1    \sqrt{1-p_1-p_2+p_1^2-p_1p_2+p_2^2}}{3(p_1-1)(p_1-p_2)}\ ,\\
        \epsilon_2&=\epsilon_1|_{p_1 \leftrightarrow p_2} \,.
\end{align}
\end{subequations}
We note that whilst $\epsilon_1$ and $\epsilon_2$ appear to have poles when $p_1 = p_2 = p$, we find that it is actually cancelled and
\begin{equation}\label{eqn:2pmix}
    \epsilon_1 = \epsilon_2 = \frac{2+p}{6(1-p)} \,.
\end{equation}
The unitarity constraints on $\epsilon_1$ and $\epsilon_2$, as in equation \eqref{eq:unit_cond}, give rise to constraints on $p_1$ and $p_2$ as
\begin{equation}\label{eqn:2gaugecond}
    p_1\geq 3 \quad \text{and}\quad p_2\geq 3\,.
\end{equation}

This constraint must be satisfied to obtain a consistent interacting SCFT with $a=c$ in the IR. Notice that the minimal value for the $p_{i}$ is $p_1=p_2=3$. In terms of the one-loop $\beta$-function for the gauge coupling, at this minimal value, we have the same coefficients as that of $SU(N_c)$ SQCD with $N_f=\frac{4}{3}N_c$ flavors, which is outside of the conformal window
\begin{align}
    \frac{3}{2}N_c \le N_f \le 3N_c \,.
\end{align}
This should not be surprising, since there exist abundant gauge theories that flow to interacting SCFTs even though the one-loop $\beta$-function coefficients lie outside of the region set by the conformal windows for SQCD \cite{Agarwal:2019crm, Agarwal:2020pol}.

When the $p_i$ satisfy the condition in equation \eqref{eqn:2gaugecond}, the central charges of the interacting SCFT in the infrared are
\begin{align}
    a=c=\frac{\left(p_1 p_2-p_1-p_2\right) \left(2Q^3-AB(A+B)\right)}{48 \left(p_1-1\right)\left(p_2-1\right)
   \left(p_1-p_2\right)^2 }\text{dim}(G)\,,
\end{align}
where we defined following quantities for convenience:
\begin{align}
A=2p_1-p_2-1\,,\quad B=-p_1+2p_2-1\,,\quad Q=\sqrt{1-p_1-p_2+p_1^2-p_1p_2+p_2^2}\,.
\end{align}
They are simplified when $p_1=p_2=p$ to
\begin{align}
    a=c=\frac{9p(p-2)}{64(p-1)}\text{dim}(G) \,. 
\end{align}

\subsection{Gauging three \texorpdfstring{$\mathcal{D}_p(G)$}{Dp(G)} theories}
\label{sec:3DpG}

Similarly to the gauging of two $\mathcal{D}_p(G)$ theories discussed in Section \ref{sec:2DpG}, superconformal field theories formed via $\mathcal{N}=1$ gauging of three $\mathcal{D}_p(G)$ are also asymptotically free for arbitrary $(p_1,p_2,p_3)$. These quivers have the following form:
\begin{align}
\begin{aligned}
    \begin{tikzpicture}
      \node[gaugeN1] (s0) {$G$};
      \node[d2] (c1) [left=0.6cm of s0] {$\mathcal{D}_{p_1}(G)$};
      \node[d2] (c2) [right=0.6cm of s0] {$\mathcal{D}_{p_2}(G)$};
      \node[d2] (c3) [above=0.4cm of s0] {$\mathcal{D}_{p_3}(G)$};
      \draw (s0.east) -- (c2.west);
      \draw (s0.west) -- (c1.east);
      \draw (s0.north) -- (c3.south);
    \end{tikzpicture} \,.
\end{aligned}
\end{align}
To check whether such a theory flows to an infrared SCFT with $a=c$, we must determine the R-charge generator of that putative SCFT.
It is technically challenging to determine the exact R-charge analytically via $a$-maximization; however, we can study the solutions in particular limits or otherwise numerically. 

We first consider the behavior when $p_2=p_1$. Then we find that the mixing coefficients $\epsilon_1$, $\epsilon_2$, and $\epsilon_3$ appearing in the infrared R-charge are
\begin{subequations}
\begin{align}
        \epsilon_1 &= \epsilon_2 = \frac{-4p_3^2+Q_0(p_1,p_3)+p_1\left(Q_1(p_1,p_3)+Q_2(p_1,p_3)^{1/2}\right)}{3 \left(p_1-1\right) Q_3(p_1,p_3)}\ ,\\
        \epsilon_3 &= -\frac{p_1 \left(-4 p_3^3+p_1 p_3^2-p_1\right)+Q_0(p_1,p_3)+p_3\left(Q_1(p_1,p_3)+2Q_2(p_1,p_3)^{1/2}\right)}{3 \left(p_3-1\right) Q_3(p_1,p_3)}\,,
\end{align}
\end{subequations}
where we have defined the polynomials $Q_j(p_1,p_3)$ as
\begin{subequations}    
\begin{align}
        Q_0(p_1,p_3)&= -2 p_1 p_3 \left(p_1-1\right) \left(p_3-1\right) ,\\
        Q_1(p_1,p_3)& = p_1^2-p_1^2p_3 +4 p_3^2\ ,\\
    \begin{split}
        Q_2(p_1,p_3)&= 9 p_1^2 \left(p_1-1\right)^4 +3\left(p_3-p_1\right) p_1 \left(p_1-1\right) \left(10 p_1^3-24 p_1^2+17 p_1-4\right) \\
        &\quad +\left(p_3-p_1\right)^2 \left(37 p_1^4-106 p_1^3+98 p_1^2-32 p_1+4\right) \\
        &\quad +4 \left(p_3-p_1\right)^3 \left(p_1-1\right) \left(5 p_1^2-7 p_1+1\right) +4 \left(p_3-p_1\right)^4 \left(p_1-1\right)^2\ ,
    \end{split}\\
    Q_3(p_1,p_3)&=-p_1^2+p_1^2p_3-4p_1p_3^2+4p_3^2\,.
\end{align}
\end{subequations}
We see that the mixing coefficients satisfy the unitarity conditions in equation \eqref{eq:unit_cond} in all cases. The central charges of the infrared SCFT are then given as
\begin{align}
    a=c=\left(\frac{(p_1p_3-p_1-p_3)Q_4(p_1,p_3)}{8(p_3-1)Q_3(p_1,p_3)}+\frac{Q_2(p_1,p_2)\left(Q_2(p_1,p_2)^{1/2}-Q_4(p_1,p_3)\right)}{12(p_1-1)(p_3-1)Q_3(p_1,p_3)^2}\right)\text{dim}(G) \,,
    \end{align}
where we additionally defined the polynomial $Q_4(p_1,p_3)$ as
\begin{align}
    Q_4(p_1,p_3)=p_1-2p_1^2+2p_3-3p_1p_3+2p_1^2p_3+2p_3^2-2p_1p_3^2 \,.
\end{align}
We further consider when all $p_i$ are identical. We find that for such a case where $p_1=p_2=p_3=p$, we get a very simple answer for the mixing coefficients:
\begin{equation}
    \epsilon_1 = \epsilon_2 = \epsilon_3 = -\frac{1}{3(p-1)}\, .
    \label{eqn:p1p2p3equals}
\end{equation}
This also simplifies the central charges to
\begin{align}
    a=c=\frac{p(2p-3)}{8(p-1)}\text{dim}(G)\,.
\end{align} 

We can also study the solutions to the $a$-maximization problem analytically in the asymptotic regime where $p_3 \gg p_1, p_2$. In this case, the mixing coefficients $\epsilon_1$, $\epsilon_2$, and $\epsilon_3$ appearing in the infrared R-charge are
\begin{subequations}
\begin{align}
        &\epsilon_1=\frac{-p_1^2(p_2-1)(p_1+p_2)+p_2Q_1(p_1,p_2)+p_1Q_2(p_1,p_2)^{1/2}}{3(p_1-1)(p_1-p_2)(p_1p_2-p_1-p_2)}+\mathcal{O}(p_3^{-1})\ ,\\
        &\epsilon_2=\epsilon_1|_{1\leftrightarrow2}\ ,\\
        &\epsilon_3=\frac{(p_1^2p_2+p_1p_2^2-p_1^2-p_2^2)Q_1(p_1,p_2)-2p_1p_2 Q_2(p_1,p_2)^{1/2}}{6(p_1-p_2)^2(p_1p_2-p_1-p_2)^2}\cdot\frac{Q_1(p_1,p_2)}{p_3}+\mathcal{O}(p_3^{-2}) \,,
\end{align}
\end{subequations}
where we have defined the polynomials $Q_1(p_1,p_2)$ and $Q_2(p_1,p_2)$ as
\begin{subequations}
\begin{align}
        Q_1(p_1,p_2)& = 2p_1p_2-p_1-p_2\ ,\\
    \begin{split}
        Q_2(p_1,p_2)&= 4p_2^2(p_2-1)^4+4(p_1-p_2)p_2(p_2-1)^3(3p_2-1) \\
        &\quad +(p_1-p_2)^2(13p_2^4-36p_2^3+33p_2^2-10p_2+1)\\
        &\quad +(p_1-p_2)^3(p_2-1)(6p_2^2-8p_2+1)+(p_1-p_2)^4(p_2-1)^2\ .
    \end{split}
\end{align}
\end{subequations}
These functions exhibit a nice simplification when $p_1=p_2 = p$ such that
\begin{align}
    Q_1(p,p)=2p(p-1)\,, \quad 
    Q_2(p,p)=4p^2(p-1)^4 \,.
\end{align}
In this special case, we find a simple expression for two of the mixing coefficients:
\begin{equation}
    \epsilon_1 = \epsilon_2 = -\frac{1}{3(p-1)} + \mathcal{O}(p_3^{-1}) \,,
\end{equation}
where we observe that the first order term is identical to the exact solution for the case of $p_1=p_2=p_3=p$ given in equation \eqref{eqn:p1p2p3equals}.
We can see that the unitarity constraints in equation \eqref{eq:unit_cond} are satisfied for all $p$. 

\begin{figure}[H]
    \centering
\begin{subfigure}{0.45\textwidth}
    \includegraphics[scale=0.45]{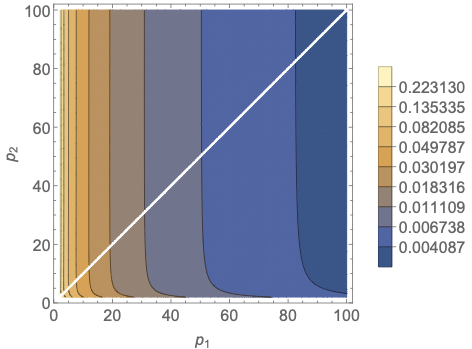}
    \caption{Contour plot of $-\epsilon_1$ on $(p_1,p_2)$ plane.}
    \label{fig:a1_large}
\end{subfigure}
\begin{subfigure}{0.45\textwidth}
    \hfill\includegraphics[scale=0.45]{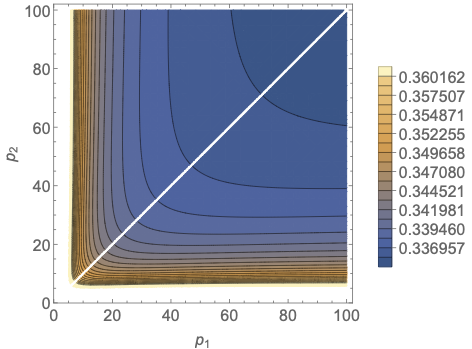}
    \caption{Contour plot of $-\epsilon_3\cdot p_3$ on $(p_1,p_2)$ plane.}
    \label{fig:a3_large}
\end{subfigure}
\caption{Contour plots of $-\epsilon_1$ and $-\epsilon_3\cdot p_3$. We see that $\epsilon_1$ lies in the range $\left(-\frac{1}{3},0\right)$, and thus $\epsilon_2$ does also. In (b) we show that $-\epsilon_3\cdot p_3$ is always positive, and since we are in a large $p_3$ limit then we can see that $\epsilon_3$ must also lie in the range $\left(-\frac{1}{3},0\right)$. 
}\label{fig:p3large}
\end{figure}

\begin{figure}[H]
    \centering
\begin{subfigure}{0.45\textwidth}
    \centering
    \includegraphics[scale=0.45]{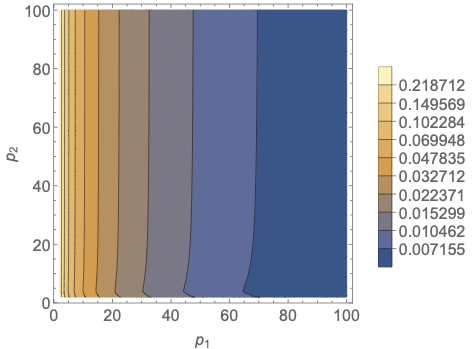}
    \caption{Contour plot of $-\epsilon_1$ for $p_3=2$.}
    \label{fig:a1_2}
\end{subfigure}
\begin{subfigure}{0.45\textwidth}
    \centering
    \includegraphics[scale=0.45]{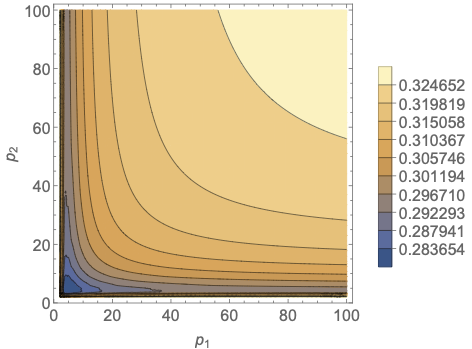}
    \caption{Contour plot of $-\epsilon_3$ for $p_3=2$.}
    \label{fig:a3_2}
\end{subfigure}\\ \vspace{0.3cm}
\begin{subfigure}{0.45\textwidth}
    \centering
    \includegraphics[scale=0.45]{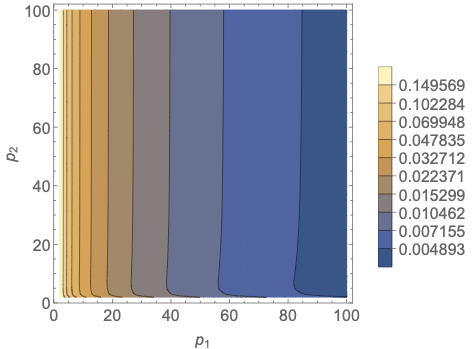}
    \caption{Contour plot of $-\epsilon_1$ for $p_3=3$.}
    \label{fig:a1_3}
\end{subfigure}
\begin{subfigure}{0.45\textwidth}
    \centering
    \includegraphics[scale=0.45]{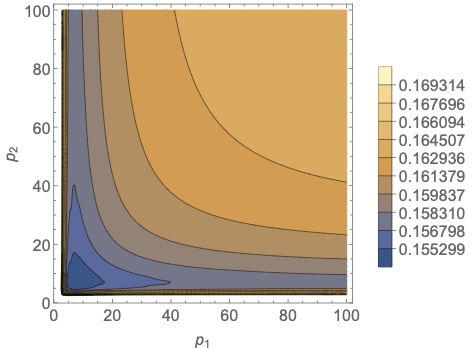}
    \caption{Contour plot of $-\epsilon_3$ for $p_3=3$.}
    \label{fig:a3_3}
\end{subfigure}\\ \vspace{0.3cm}
\begin{subfigure}{0.45\textwidth}
    \centering
    \includegraphics[scale=0.45]{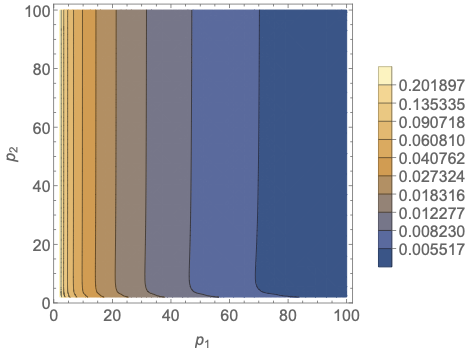}
    \caption{Contour plot of $-\epsilon_1$ for $p_3=4$.}
    \label{fig:a1_4}
\end{subfigure}
\begin{subfigure}{0.45\textwidth}
    \centering
    \includegraphics[scale=0.45]{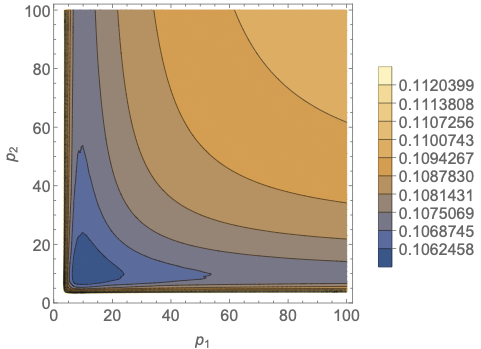}
    \caption{Contour plot of $-\epsilon_3$ for $p_3=4$.}
    \label{fig:a3_4}
\end{subfigure}
\caption{Contours plot of $\epsilon_1$ and $\epsilon_3$ in the $(p_1,p_2)$ plane for $p_3=2,3,4$. They all satisfy the unitarity condition in equation \eqref{eq:unit_cond}.}\label{fig:smallp3}
\end{figure}

For general cases where $p_1 \neq p_2$ we do a numerical check of the values of the mixing parameters $\epsilon_i$. We can see from the contour plots on the $(p_1, p_2)$-plane in Figure \ref{fig:p3large} that the $\epsilon_i$ are in the range $-\frac{1}{3} < \epsilon_i < 0$, when $p_3$ is large, which is sufficient for them to flow to an SCFT with $a = c$. In addition, we checked numerically that the unitarity condition in equation \eqref{eq:unit_cond} is satisfied for small values of $p_3$, specifically for $p_3 = 2, 3, 4$, as can be seen in Figure \ref{fig:smallp3}. In both the small and the large $p_3$ regime, the unitarity condition is always satisfied in the infrared limit of three $\mathcal{D}_{p_i}(G)$ theories gauged together. Considering the monotonicity of the $\epsilon_i$ along $p_1$ and $p_2$ in both regimes, we expect they will monotonically increase along with the increasing $p_3$ and thus that the unitarity condition is always satisfied. In short, every gluing $(p_1, p_2, p_3)$ leads to an infrared SCFT with $a = c$.

\subsection{Gauging four \texorpdfstring{$\mathcal{D}_p(G)$}{Dp(G)} theories}
\label{sec:4DpG}

\begin{figure}[H]
    \centering
\begin{subfigure}{0.48\textwidth}
    \includegraphics[scale=0.45]{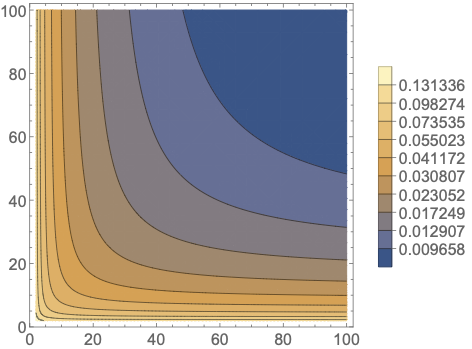}
    \caption{Contour plot of $-\epsilon_1$ on the $(p_3,p_4)$-plane.}
\end{subfigure}
\begin{subfigure}{0.48\textwidth}
    \hfill \includegraphics[scale=0.45]{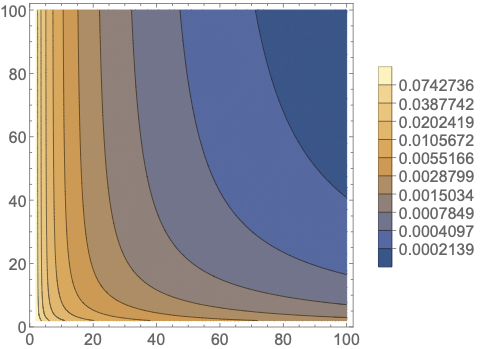}
    \caption{Contour plot of $-\epsilon_3$ on the $(p_3,p_4)$-plane.}
\end{subfigure}
\caption{Contour plot of $-\epsilon_1$ and $-\epsilon_3$ for $(p_1,p_2,p_3,p_4)=(2,2,p_3,p_4)$. Every R-symmetry mixing coefficient $\epsilon_i$ satisfies the unitarity condition in equation \eqref{eq:unit_cond}.}
    \label{fig:a_4g_large}
\end{figure}

\begin{figure}[H]
    \centering
\begin{subfigure}{0.24\textwidth}
    \centering
    \includegraphics[scale=0.31]{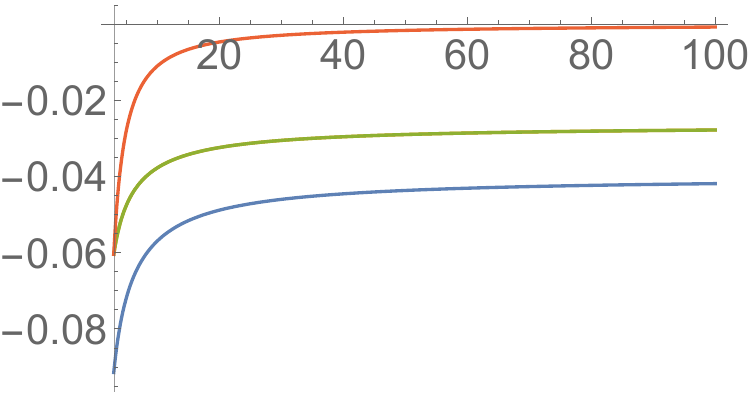}
    \caption{$(2,3,3,p_4)$}
\end{subfigure}
\begin{subfigure}{0.24\textwidth}
    \centering
    \includegraphics[scale=0.31]{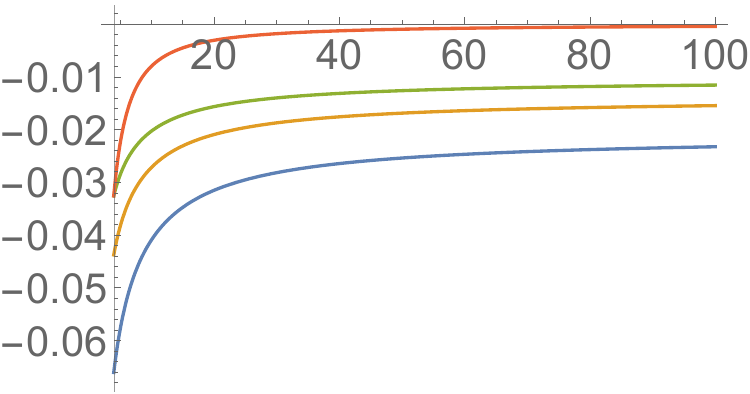}
    \caption{$(2,3,4,p_4)$}
\end{subfigure}
\begin{subfigure}{0.24\textwidth}
    \centering
    \includegraphics[scale=0.31]{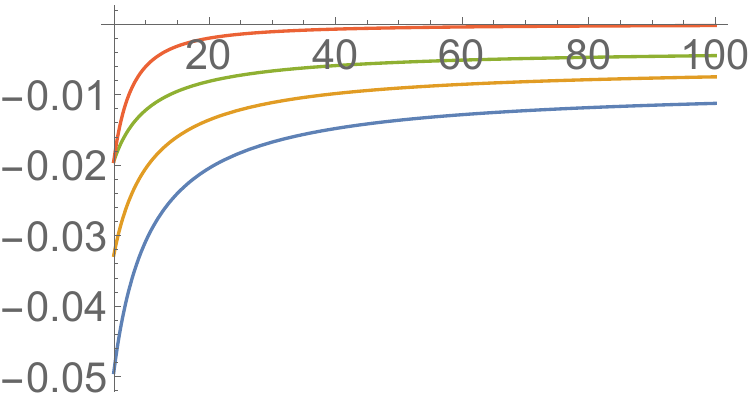}
    \caption{$(2,3,5,p_4)$}
\end{subfigure}
\begin{subfigure}{0.24\textwidth}
    \centering
    \includegraphics[scale=0.31]{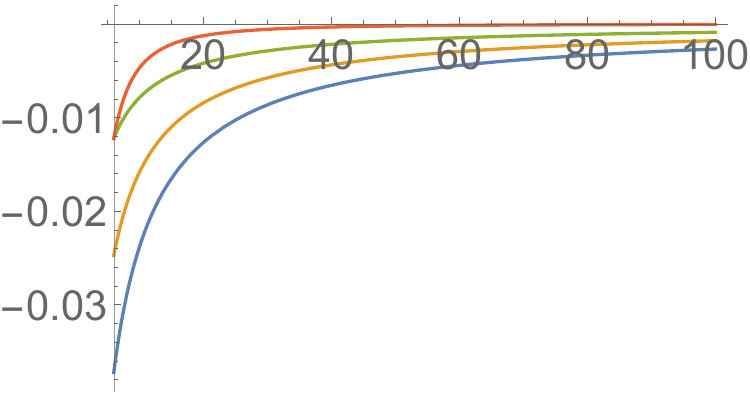}
    \caption{$(2,3,6,p_4)$}
\end{subfigure}\\[6pt]
    \centering
\begin{subfigure}{0.24\textwidth}
    \centering
    \includegraphics[scale=0.31]{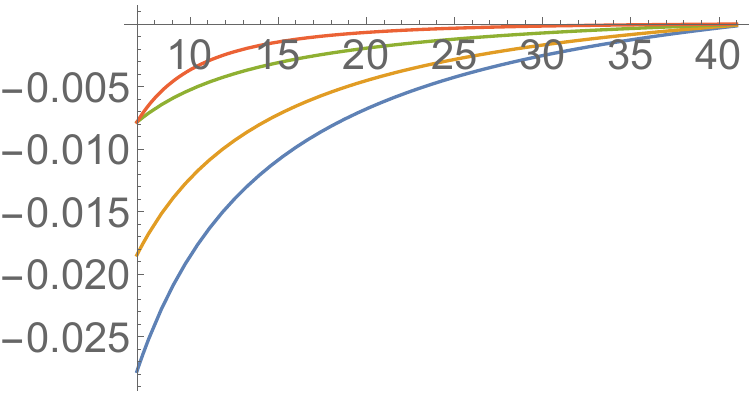}
    \caption{$(2,3,7,p_4)$}
\end{subfigure}
\begin{subfigure}{0.24\textwidth}
    \centering
    \includegraphics[scale=0.31]{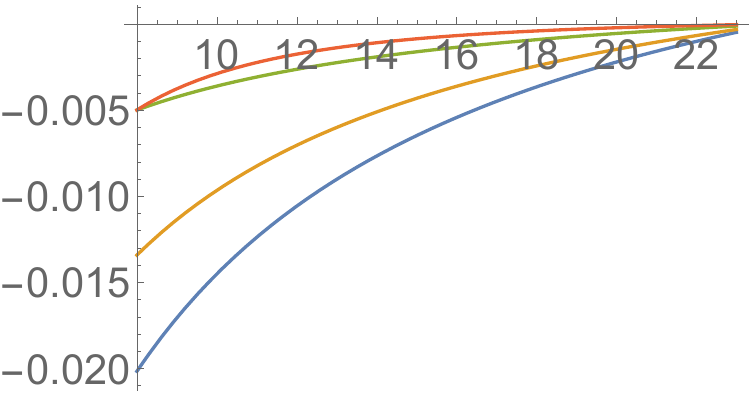}
    \caption{$(2,3,8,p_4)$}
\end{subfigure}
\begin{subfigure}{0.24\textwidth}
    \centering
    \includegraphics[scale=0.31]{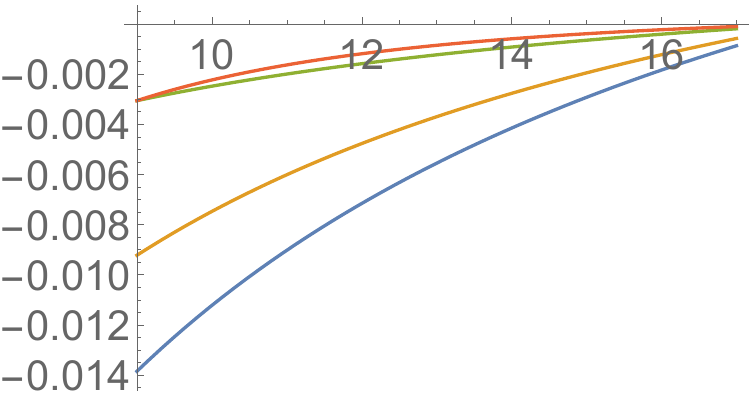}
    \caption{$(2,3,9,p_4)$}
\end{subfigure}
\begin{subfigure}{0.24\textwidth}
    \centering
    \includegraphics[scale=0.31]{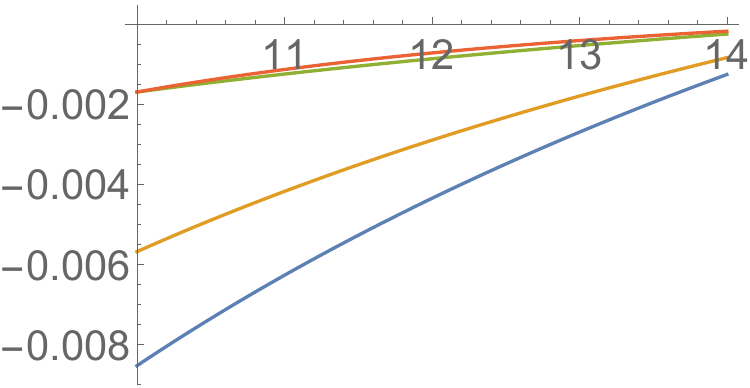}
    \caption{$(2,3,10,p_4)$}
\end{subfigure}\\[6pt]
\begin{subfigure}{0.24\textwidth}
    \centering
    \includegraphics[scale=0.31]{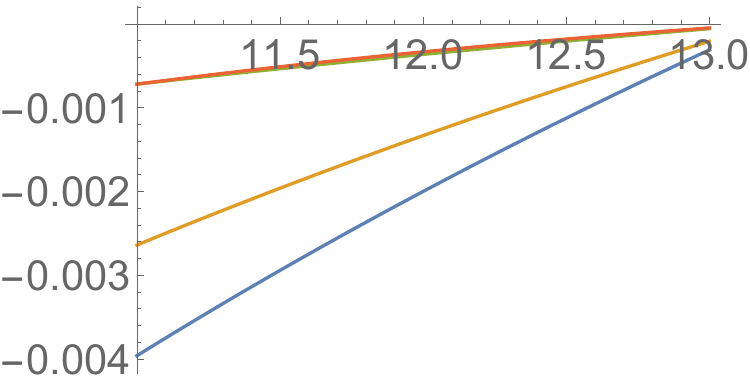}
    \caption{$(2,3,11,p_4)$}
\end{subfigure}
\begin{subfigure}{0.24\textwidth}
    \centering
    \includegraphics[scale=0.31]{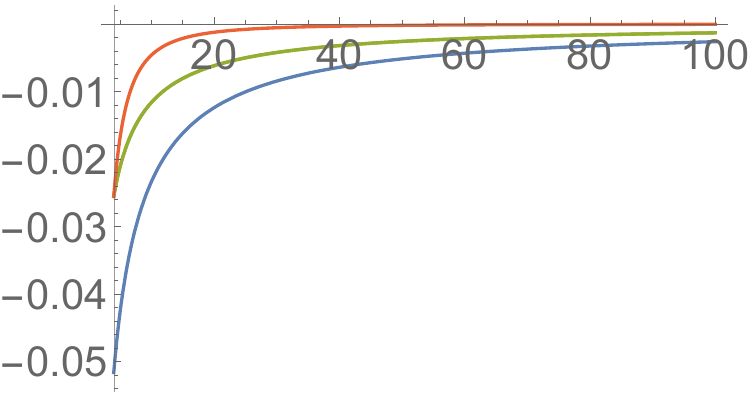}
    \caption{$(2,4,4,p_4)$}
\end{subfigure}
\begin{subfigure}{0.24\textwidth}
    \centering
    \includegraphics[scale=0.31]{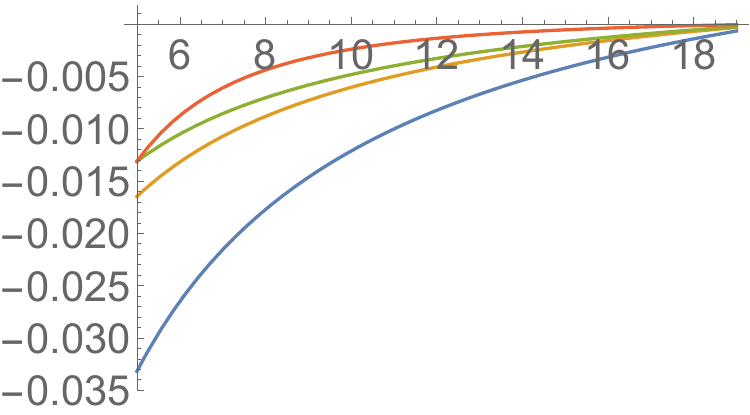}
    \caption{$(2,4,5,p_4)$}
\end{subfigure}
\begin{subfigure}{0.24\textwidth}
    \centering
    \includegraphics[scale=0.31]{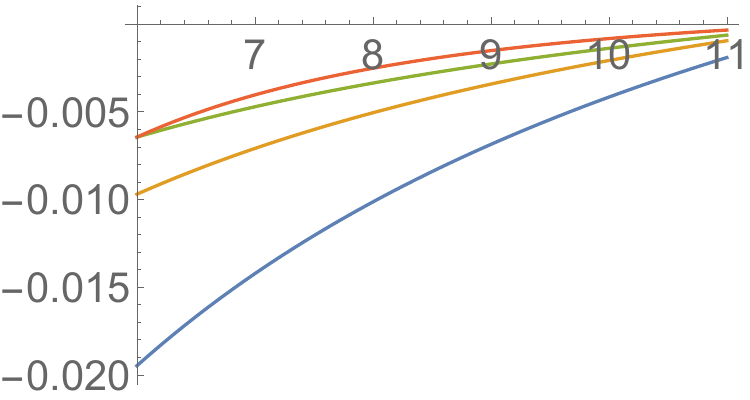}
    \caption{$(2,4,6,p_4)$}
\end{subfigure}\\[6pt]
\begin{subfigure}{0.24\textwidth}
    \centering
    \includegraphics[scale=0.31]{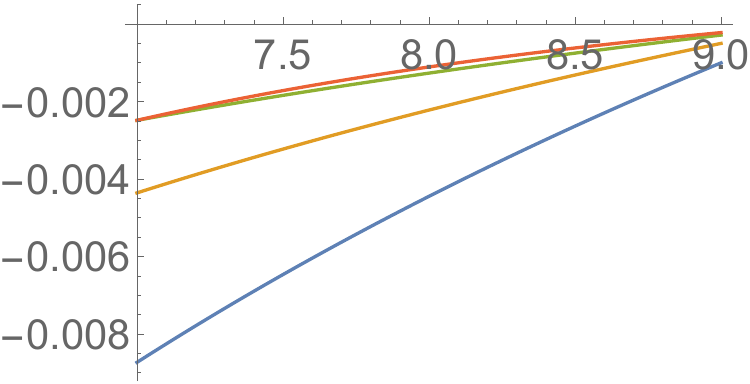}
    \caption{$(2,4,7,p_4)$}
\end{subfigure}
\begin{subfigure}{0.24\textwidth}
    \centering
    \includegraphics[scale=0.31]{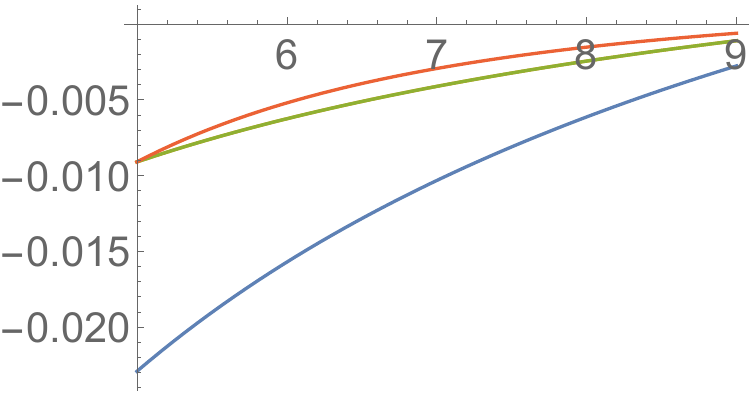}
    \caption{$(2,5,5,p_4)$}
\end{subfigure}
\begin{subfigure}{0.24\textwidth}
    \centering
    \includegraphics[scale=0.31]{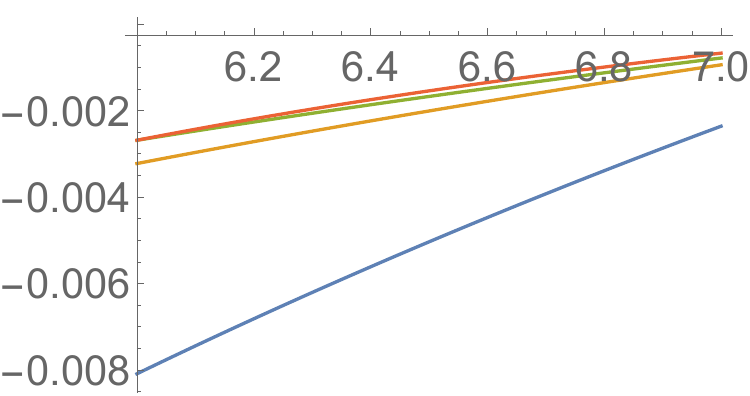}
    \caption{$(2,5,6,p_4)$}
\end{subfigure}
\begin{subfigure}{0.24\textwidth}
    \centering
    \includegraphics[scale=0.31]{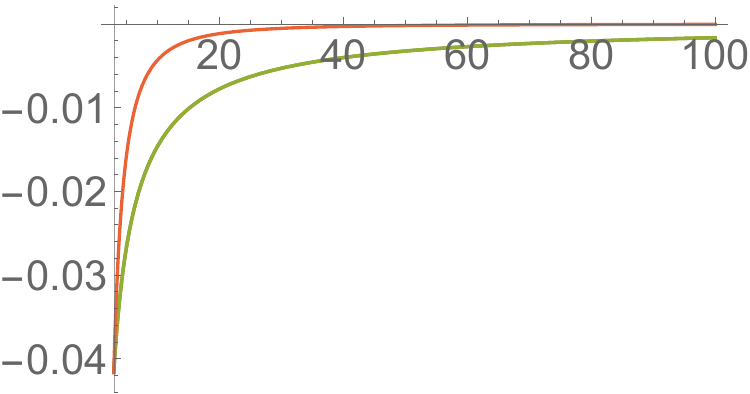}
    \caption{$(3,3,3,p_4)$}
\end{subfigure}\\[6pt]
    \centering
\begin{subfigure}{0.3\textwidth}
    \centering
    \includegraphics[scale=0.37]{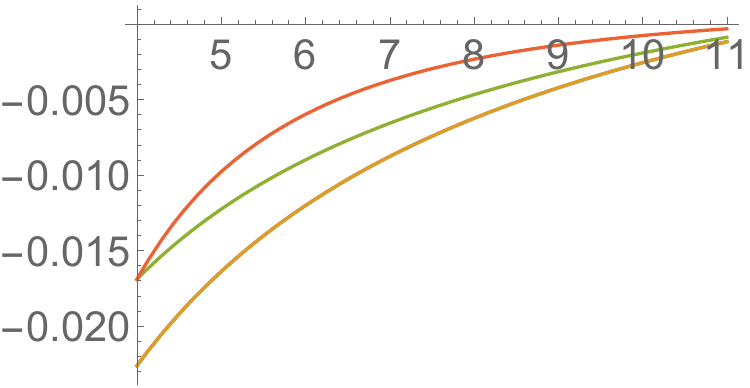}
    \caption{$(3,3,4,p_4)$}
\end{subfigure}
\begin{subfigure}{0.3\textwidth}
    \centering
    \includegraphics[scale=0.37]{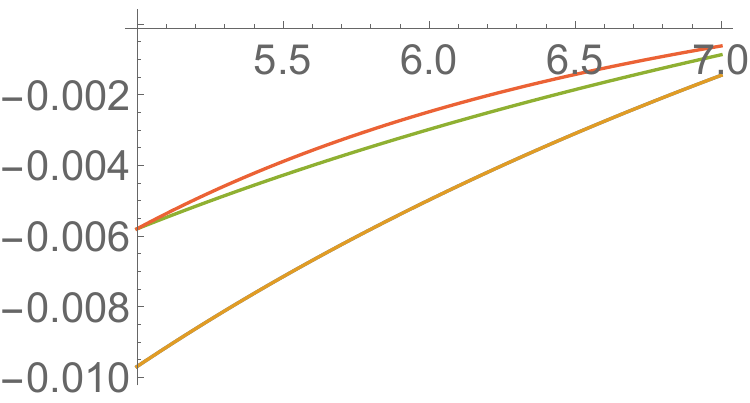}
    \caption{$(3,3,5,p_4)$}
\end{subfigure}
\begin{subfigure}{0.3\textwidth}
    \centering
    \includegraphics[scale=0.37]{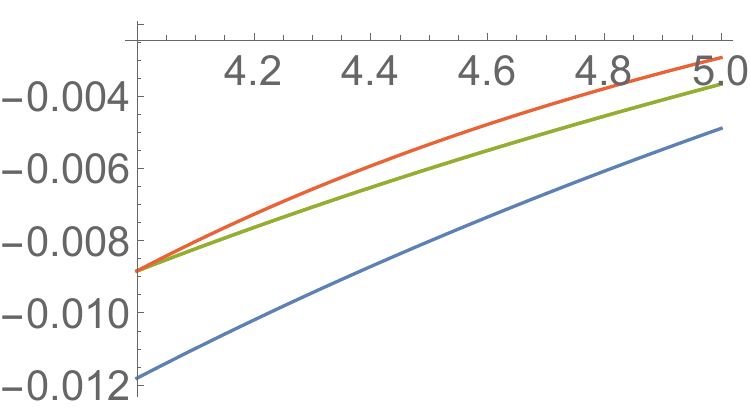}
    \caption{$(3,4,4,p_4)$}
\end{subfigure}
\caption{Plot of $\epsilon_{1,2,3,4}$ in gluing four $\mathcal{D}_{p_i}(G)$ theories with all $(p_1,p_2,p_3,p_4)$ listed in Table \ref{tbl:asympfreep}, except first case, $(2,2,p_3,p_4)$, which is numerically studied in Figure \ref{fig:a_4g_large}. All the coefficients are in the range prescribed in equation \eqref{eq:unit_cond}.}
\label{fig:a_4g}
\end{figure}

We turn now to the case where we gauge together four $\mathcal{D}_p(G)$ theories. These 4d $\mathcal{N}=1$ SCFTs can be written as the infrared fixed points of the diagonal gauging, which can be depicted as
\begin{align}
\begin{aligned}
    \begin{tikzpicture}
      \node[gaugeN1] (s0) {$G$};
      \node[d2] (c1) [left=1.0cm of s0] {$\mathcal{D}_{p_1}(G)$};
      \node[d2] (c2) [right=1.0cm of s0] {$\mathcal{D}_{p_3}(G)$};
      \node[d2] (c3) [above=0.4cm of s0] {$\mathcal{D}_{p_4}(G)$};
      \node[d2] (c4) [below=0.4cm of s0] {$\mathcal{D}_{p_2}(G)$};
      \draw (s0.east) -- (c2.west);
      \draw (s0.west) -- (c1.east);
      \draw (s0.north) -- (c3.south);
      \draw (s0.south) -- (c4.north);
    \end{tikzpicture}
\end{aligned} \,.
\end{align}
All gaugings which satisfy the asymptotic freedom condition are listed in Table \ref{tbl:asympfreep}, and each of these $(p_1, p_2, p_3, p_4)$ satisfy the unitarity conditions in equation \eqref{eq:unit_cond}. Each case has been checked numerically. In Figure \ref{fig:a_4g_large}, we show the contour plots for the $\epsilon_i$ on the $(p_3, p_4)$-plane for the gauging $(2,2, p_3,p_4)$. In Figure \ref{fig:a_4g}, we write the value of $\epsilon_i$ against the value of $p_4$ for each of the other gaugings in Table \ref{tbl:asympfreep}. We can observe that the unitarity condition is always satisfied for all possible gaugings either explicitly or via the asymptotic behavior when $p_3$ or $p_4$ is unbounded. Similar to the case where three $\mathcal{D}_p(G)$ theories are gauged together, all gaugings of four $\mathcal{D}_p(G)$ theories lead to interacting superconformal field theories with $a = c$ in the infrared.\\

\subsection{Gauging five \texorpdfstring{$\mathcal{D}_p(G)$}{Dp(G)} theories}
\label{sec:5DpG}

Finally, we turn our attention to the maximal number of $\mathcal{D}_p(G)$ theories that can be gauged together with an asymptotically free coupling: five $\mathcal{D}_p(G)$. There are only four classes of theories with five $\mathcal{D}_p(G)$ theories glued together in such a way, as can be seen from Table \ref{tbl:asympfreep}. These give rise to the quivers
\begin{align}
\begin{aligned}
    \begin{tikzpicture}
      \node[gaugeN1] (s0) {$G$};
      \node[d2] (c1) [left=1.0cm of s0] {$\mathcal{D}_{p_3}(G)$};
      \node[d2] (c2) [right=1.0cm of s0] {$\mathcal{D}_{p_5}(G)$};
      \node[d2] (c3) [above=0.4cm of s0] {$\mathcal{D}_{p_4}(G)$};
      \node[d2] (c4) [below right=0.4cm of c1] {$\mathcal{D}_{p_2}(G)$};
      \node[d2] (c5) [below left=0.4cm of c2] {$\mathcal{D}_{p_1}(G)$};
      \draw (s0.east) -- (c2.west);
      \draw (s0.west) -- (c1.east);
      \draw (s0.north) -- (c3.south);
      \draw (s0.225) -- (c4.45);
      \draw (s0.315) -- (c5.135);
    \end{tikzpicture} \,.
\end{aligned}
\end{align}
We can check analytically that the unitarity bound in equation \eqref{eq:unit_cond} holds for each case. The R-symmetry mixing parameters after $a$-maximization are found to be
\begin{subequations}\label{eqn:swiper}
\begin{align}
    \begin{split}
        (2,2,2,2,p)\ :& \  \epsilon_{1,2,3,4}=\frac{2p^2-2-Q(p)^{1/2}}{3(4p^2-p+1)}\ ,\epsilon_5=\frac{-4p^3-4p^2+5p-1+2pQ(p)^{1/2}}{3(p-1)(4p^2-p+1)}\ ,\\
        & \ Q(p)=4p^4+8p^3-14p^2+3p+3\ ,  
    \end{split}\\
    \begin{split}
        (2,2,2,3,3)\ :& \ \epsilon_{1,2,3}=\frac{1}{147}(163-4\sqrt{1731})\sim -0.0232731\ ,\\
        &\ \epsilon_{4,5}=\frac{1}{294}(-379+9\sqrt{1731})\sim -0.0154844\ ,
    \end{split}\\
        (2,2,2,3,4)\ :& \ \epsilon_{1,2,3}\sim -0.0119828\ ,\quad \epsilon_4\sim -0.00798042\ ,\quad \epsilon_5\sim-0.00597775\ ,\\
        (2,2,2,3,5)\ :& \ \epsilon_{1,2,3}\sim -0.00490809\ ,\quad \epsilon_4\sim-0.00327071\ ,\quad \epsilon_5\sim-0.00196062 \,.
    \label{eq:pset_5}
\end{align}
\end{subequations}
In the first case it is easy to check that the $\epsilon_i$ satisfy the unitarity condition in equation \eqref{eq:unit_cond} for any value of $p$. Thus all the cases of equation \eqref{eqn:swiper} flow in the infrared to interacting SCFTs with identical central charges. We list the values of those central charges, in each case, in Table \ref{tab:cc5DpG}.

\begin{table}[H]
\begin{threeparttable}
\centering
\renewcommand{\arraystretch}{1.5}
    $
    \begin{array}{cc}
    \toprule
    (p_1,p_2,p_3,p_4,p_5) & a=c \\\midrule\\[-19pt]
    (2,2,2,2,p)\ &\ \left(\dfrac{p(2p^2-p+3)(8p^3+28p^2-55p+15)+2Q(p)^{3/2}}{24(p-1)(4p^2-p+1)^2}\right)\text{dim}(G)\ \\[15pt]
    (2,2,2,3,3)\ &\ \left(\dfrac{3(-19458+577\sqrt{1731})}{19208}\right)\text{dim}(G)\sim0.710368\ \text{dim}(G)\ \\[13pt]
    (2,2,2,3,4)\ &\ \sim 0.793031\ \text{dim}(G)\ \\
    (2,2,2,3,5)\ &\ \sim 0.875633\ \text{dim}(G)\ \\\bottomrule
    \end{array}
    $
\end{threeparttable}
    \caption{Central charges $a=c$ of the infrared SCFTs obtained by gauging together five $\mathcal{D}_{p_i}(G)$ theories with an $\mathcal{N}=1$ vector multiplet.}
    \label{tab:cc5DpG}
\end{table}

\subsection{Gauging six \texorpdfstring{$\mathcal{D}_2(G)$}{D2(G)} theories: conformal gauging}
\label{sec:6DpG}

Throughout Section \ref{sec:Neq1gluings} thus far, we mainly focused on asymptotically-free $(\mathcal{N}=1)$-gaugings of $\mathcal{D}_{p_i}(G)$ theories, and the $a$-maximization procedure that must be carried out to determine the superconformal R-symmetry. In addition to the asymptotically-free gaugings, as shown in Table \ref{tbl:asympfreep}, there are collections of $p_i$ that can be combined via conformal gaugings; hence, the theory does not flow. We list the conformal gaugings in Table \ref{tbl:confgaugep}. We do not study in detail the conformal gaugings in this paper, as they are not \emph{a priori} guaranteed to have the exactly marginal operators that are required for the theories to be interacting SCFTs with $a = c$. However, in this subsection, we highlight a particular example. We consider the theory formed by $\mathcal{N}=1$ gauging of six copies of $\mathcal{D}_2(G)$, as depicted in the quiver:
\begin{align}
\begin{aligned}
    \begin{tikzpicture}
      \node[gaugeN1] (s0) {$G$};
      \node[d2] (c1) [left=1.0cm of s0] {$\mathcal{D}_{2}(G)$};
      \node[d2] (c2) [right=1.0cm of s0] {$\mathcal{D}_{2}(G)$};
      \node[d2] (c3) [above right=0.4cm of c1] {$\mathcal{D}_{2}(G)$};
      \node[d2] (c0) [above left=0.4cm of c2] {$\mathcal{D}_{2}(G)$};
      \node[d2] (c4) [below right=0.4cm of c1] {$\mathcal{D}_{2}(G)$};
      \node[d2] (c5) [below left=0.4cm of c2] {$\mathcal{D}_{2}(G)$};
      \draw (s0.east) -- (c2.west);
      \draw (s0.west) -- (c1.east);
      \draw (s0.45) -- (c0.225);
      \draw (s0.135) -- (c3.315);
      \draw (s0.225) -- (c4.45);
      \draw (s0.315) -- (c5.135);
    \end{tikzpicture} \,.
\end{aligned}
\end{align}
Each $\mathcal{D}_2(G)$ theory has a Coulomb branch operator of lowest dimension $u_i^{0}$, whose dimension is given by
\begin{equation}
    \Delta(u_i^{0}) = \frac{p_i+1}{p_i} = \frac{3}{2} \,.
\end{equation}
Henceforth we see that the gauged theory has $21$ marginal operators formed from the product of these Coulomb branch operators: 
\begin{align}
    u_i^{0}u_j^{0}\, .
\end{align}
Furthermore, there are five $U(1)$ flavor currents formed from non-anomalous combinations of the six $U(1)$ symmetries, which is from the $\mathcal{N}=2$ R-symmetry of each $\mathcal{D}_2(G)$ theory. These symmetries are generically broken by the aforementioned marginal operators, so that the broken currents combine with them to become marginally irrelevant \cite{Leigh:1995ep, Green:2010da}. Therefore we are left with sixteen exactly marginal operators which span a conformal manifold. All other combinations of more than one Coulomb branch operator are irrelevant. Due to the presence of the exactly marginal operators, we can see that this gauging provides non-trivial SCFTs with $a = c$. Indeed, we find
\begin{equation}
    a = c = \frac{5}{8} \operatorname{dim}(G) \,.
\end{equation}

\section{\texorpdfstring{\boldmath{$\mathcal{N}=1$}}{N=1} gluing with adjoint chirals}
\label{sec:withchirals}

In Section \ref{sec:Neq1gluings}, we study how to glue together several $\mathcal{D}_p(G)$ SCFTs via $(\mathcal{N}=1)$-gauging of the diagonal flavor symmetry $G$. We introduce a new $G$-vector multiplet and couple it to the flavor current, as in equation \eqref{eqn:supcup}. In this section, we consider a more general gluing, where, in addition to the $G$-vector multiplet, we include some number of chiral multiplets charged under $G$. Let us introduce $n$ chiral multiplets in the representation $\mathbf{R}_{\ell=1, \ldots, n}$ of $G$ and with R-charge $R_\ell$. Then the condition for the R-symmetry to be non-anomalous is that
\begin{align}
\begin{split}
    0=\trace RGG &= h^\vee_G+\sum_i\left(\left(\frac{1}{3}-\epsilon_i\right)\trace_iR_{\mathcal{N}=2}GG+\left(\frac{4}{3}+2\epsilon_i\right)\trace_iI_3GG\right) \cr & \qquad\quad +\sum_{\ell=1}^n(R_\ell-1)I(\mathbf{R_\ell})\\
    &= h^\vee_G+\sum_i\left(\frac{1}{3}-\epsilon_i\right)\left(-\frac{p_i-1}{p_i}h^\vee_G\right)+\sum_\ell(R_\ell-1)I(\mathbf{R_\ell}) \,,
\end{split}
\end{align}
where $I(\mathbf{R_\ell})$ is the Dynkin index of the representation $\mathbf{R_\ell}$ in $G$, and $\trace_i$ refers to the anomaly coefficients for the $\mathcal{D}_{p_i}$ theories that we gauge. 

On the other hand, the difference between the two central charges $(a-c)$ is given by
\begin{align}\label{eq:trR_2}
    \begin{split}
    \trace R&=16(a-c)\\
    &=\textrm{dim}(G)+\sum_i\left(\left(\frac{1}{3}-\epsilon_i\right) \trace_i R_{\mathcal{N}=2}+\left(\frac{4}{3}+2\epsilon_i\right)\trace_iI_3\right)+\sum_\ell (R_\ell-1)\textrm{dim}(\mathbf{R_\ell})\\
    &=\textrm{dim}(G)+\sum_i\left(\frac{1}{3}-\epsilon_i\right)48(a_i-c_i)+\sum_\ell(R_\ell-1)\textrm{dim}(\mathbf{R_\ell})\ .
    \end{split}
\end{align}
We find that the RGG-anomaly cancellation guarantees $a=c$, as before, when
\begin{align}\label{eqn:acadjineq}
    \frac{\textrm{dim}(G)}{h^\vee_G}=\frac{48(a_i-c_i)}{-\frac{p_i-1}{p_i}h^\vee_G}=\frac{\textrm{dim}(\mathbf{R_\ell})}{I(\mathbf{R_\ell})} \,,
\end{align}
for each $\mathcal{D}_{p_i}(G)$ theory and each chiral matter representation $\bm{R_\ell}$. Then the RGG-anomaly cancellation directly leads to $a=c$ via
\begin{align}
    \begin{split}
        a-c=\frac{\textrm{dim}(G)}{16h^\vee_G}\trace RGG=0\ ,
    \end{split}
\end{align}
as was the case in the previous section. The first equality in equation \eqref{eqn:acadjineq} holds when $p_i$ and $h^\vee_G$ are coprime. The second equality holds if the matter is in the adjoint representation. Thus $\mathcal{D}_p(G)$ theories glued together with additional adjoint chiral multiplets may flow to interacting infrared SCFTs with $a=c$. 

Asymptotic freedom or vanishing of the one-loop $\beta$-function is required to have an interacting SCFT. It restricts the possible set of gaugings to those satisfying
\begin{align}
    \begin{split}
        \sum_i\frac{p_i-1}{p_i}\leq 3-n_a \,,
    \end{split}
    \label{eq:asymp_free_adj}
\end{align}
where $n_a$ is the number of adjoint chiral multiplets. Clearly, a theory satisfying equation \eqref{eq:asymp_free_adj} can have only 
\begin{align}
    n_a = 0, 1, 2.
\end{align}
Let us investigate these cases.\footnote{There is also a theory with $n_a = 3$, where all $p_i = 1$. This is simply the Lagrangian gauge theory with three adjoint chiral multiplets; which sits on the conformal manifold containing $\mathcal{N}=4$ super  Yang--Mills. For obvious reasons, we will not discuss this case further.} We already discussed the cases with $n_a=0$ in Section \ref{sec:Neq1gluings}. When $n_a=1$, we have theories which may be connected to the $\mathcal{N}=2$ gaugings through superpotential deformations, and we study these further in Sections \ref{sec:N2pot} and \ref{sec:massdef}. Including two adjoint chiral multiplets (i.e. $n_a=2$) leads to an asymptotically free theory only when one has a single $\mathcal{D}_p(G)$ theory being gauged, for any value of $p$. A theory with vanishing one-loop $\beta$-function is obtained only when the diagonal of the flavor symmetry of two $\mathcal{D}_2(G)$ theories is gauged. We explore the $n_a=2$ cases in Sections \ref{sec:lagrangian} and \ref{sec:2adjoints}.

\subsection{\texorpdfstring{$\mathcal{N}=1$}{N=1} gluing with one adjoint chiral}
\label{sec:N2pot}

In this subsection, we consider several $\mathcal{D}_p(G)$ theories glued by $\mathcal{N}=1$ gauging, together with one chiral multiplet $\phi$ in the adjoint representation of $G$. There are two possibilities, the introduced gauge coupling can either be asymptotically free or it can have vanishing one-loop $\beta$-function. In the latter case, the matter content is the same as with the theories formed via $\mathcal{N}=2$ conformal gauging, and the supersymmetry enhances to $\mathcal{N}=2$ if a superpotential term
\begin{align}
    \begin{split}
         W=\sum_i \trace \mu_i \phi \,,
    \end{split}
    \label{eq:N=2suppot}
\end{align}
is turned on. The explicit sets of $p_i$ that satisfy the condition written in equation \eqref{eq:asymp_free_adj}, and are thus asymptotically free or have vanishing one-loop $\beta$-function are given by
\begin{align}\label{eqn:1adjg}
    \begin{split}
       p_i = (p_1), \, (p_1, p_2),\ (2, 2, p_3),\ (2,3,\leq 6),\ (2,4,4),\ (3,3,3),\ (2,2,2,2) \,,
    \end{split}
\end{align}
where the $p_i$ are given in ascending order as before, and when we write a $p_i$ in the tuple then any positive integer $\geq 2$ satisfies equation \eqref{eq:asymp_free_adj}.

The first case to consider is when we gauge a single $\mathcal{D}_p(G)$ together with one additional chiral multiplet. In this case, the R-charge of the adjoint chiral multiplet $\phi$ is fixed by the cancellation of the R-gauge-gauge anomaly:
\begin{equation}
    R_\phi = \left(\frac{1}{3} - \epsilon\right)\left(\frac{p-1}{p}\right) \,.
\end{equation}
Maximizing $a$ with respect to $\epsilon$ leads to
\begin{equation}
    \epsilon = \frac{1 - 2p^2 + p \sqrt{4p^2 -6p + 3}}{6p^2 -9p + 3} \,.
\end{equation}
Thus, at the infrared fixed point we have the following R-charges of the fields
\begin{equation}
    \begin{aligned}
        \frac{1}{9}(5 - \sqrt{7}) \leq R_\phi &= \frac{4p-3-\sqrt{4 p^2-6 p+3}}{6p-3} < \frac{1}{3} \,,\\
        \frac{2}{9}(2\sqrt{7}-1) \leq R_\mu &= \frac{4p^2 - 12p + 6 + 2p \sqrt{4 p^2-6 p+3}}{6 p^2-9 p+3} < \frac{4}{3} \,.
    \end{aligned}
\end{equation}
We can see that $\trace \phi^2$ violates the unitarity bound, which signals existence of an accidental $U(1)$ flavor symmetry that acts on the decoupled free operator $\trace \phi^2$ \cite{Kutasov:2003iy}. This invalidates our $a$-maximization analysis and the previous argument in Section \ref{sec:acgauging} that the interacting SCFT has $a=c$ does not apply. 

One convenient way to remove such a decoupled operator is by adding a flipper-field $M$ and the corresponding superpotential term \cite{Barnes:2004jj, Benvenuti:2017lle, Maruyoshi:2018nod}
\begin{align}
    \Delta W=M\trace \phi^2 \,.
\end{align}
This introduces a $U(1)$ flavor symmetry, under which only the operator that gets decoupled is charged, which subsequently mixes with the superconformal R-symmetry during the flow into the infrared. After removing the $\trace \phi^2$ operator via flipping and flowing into the infrared, one finds that the superconformal R-charges in the interacting sector. For example, when $G=SU(3)$, they are  \begin{equation}
    \begin{aligned}
        \frac{1}{48}(27 - \sqrt{217}) \leq R_\phi &= \frac{8p^2-p-3-\sqrt{16p^4-15p^2+6p+9}}{12p^2} < \frac{1}{3} \,,\\
        \frac{1}{12}(\sqrt{217}-3) \leq R_\mu &= \frac{4 p^2-11 p+3+\sqrt{16 p^4-15 p^2+6 p+9}}{6 p(p-1) } < \frac{4}{3} \,, \\
        R_M &= 2 - 2R_\phi \,.
    \end{aligned}
\end{equation}
Thus, it is straightforward to determine that the two central charges are given by
\begin{equation}\label{eqn:acflip}
    a = a_0 \operatorname{dim}(G) - a_1 \,, \qquad c = c_0 \operatorname{dim}(G) - c_1 \,,
\end{equation}
where the coefficients of the $\operatorname{dim}(G)$ terms in both central charges are equal: $a_0 = c_0$. More specifically, the $p$-dependent constants appearing in equation \eqref{eqn:acflip} are
\begin{subequations}
\begin{align}
\begin{split}
    a_0 &= c_0 = \frac{(p+1) (2 p-3) \left(32 p^6+8 p^5+6 p^4-19 p^3+11 p^2+3 p-9\right)}{1536 (p-1) p^6}\\
    &\qquad\qquad\quad+\frac{(p+1) Q(p) \left(16 p^5-20 p^4+12 p^3+3 p^2-12 p+9\right)}{1536 (p-1) p^6} \,, \\
\end{split}\\
    a_1 &= \frac{(p+1)^2 (2 p-3)(10 p^3-11 p^2-3 p+9)}{192 p^6}-\frac{(4 p^4-3 p^3-12 p^2+6 p+9) Q(p)}{192 p^6} ,\\
    c_1 &= \frac{(p+1)(2p-3)(8 p^4-p^3-14 p^2+6 p+9)}{192 p^6}-\frac{(p-3)(2 p^3+3p^2- 3p-3)Q(p)}{192 p^6} ,
\end{align}
where we have again defined a useful function
\begin{align}
    Q(p) = \sqrt{16 p^4-15 p^2+6 p+9}\, .
\end{align}
\end{subequations}
Here the $a_1$ and $c_1$ are the central charges of the flipper-field $M$ that removes the decoupling $\text{Tr}\phi^2$.
Thus, (the interacting sector of) the infrared SCFT obtained from gauging a single $\mathcal{D}_p(G)$ theory, together with an additional chiral multiplet in the adjoint representation of $G$, does not have $a = c$. One can do the same analysis with other $G$ and find that the same happens for other gauge groups as well.

We now study the remaining combinations of $\{p_i\}$ in equation \eqref{eqn:1adjg} that can be gauged together, with an extra adjoint chiral, in such a way that the gauge coupling is either asymptotically-free or conformal. We find that, in all cases in equation \eqref{eqn:1adjg} where we gauge together more than one $\mathcal{D}_p(G)$, if each $p_i$ satisfies
\begin{align}
    \gcd(p_i, h_G^\vee) = 1 \,,
\end{align}
then we obtain an infrared SCFT with $a = c$.

\begin{figure}[H]
    \centering
\begin{subfigure}{0.45\textwidth}
    \centering
    \includegraphics[scale=0.56]{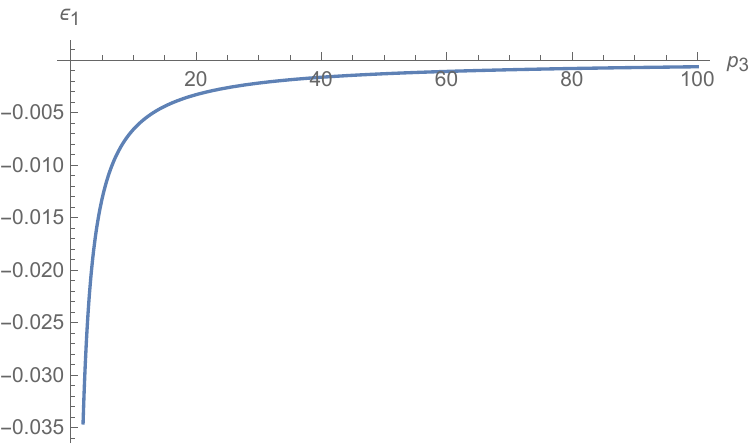}
    \caption{Plot of $\epsilon_1$ against $p_3$ for the gauging $(2,2,p_3)$ with one adjoint chiral.}
\end{subfigure}\qquad
\begin{subfigure}{0.45\textwidth}
    \centering
    \includegraphics[scale=0.56]{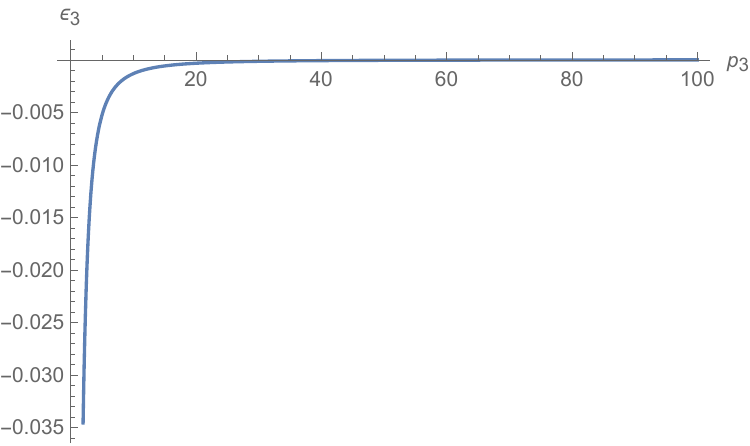}
    \caption{Plot of $\epsilon_3$ against $p_3$ for the gauging $(2,2,p_3)$ with one adjoint chiral.}
\end{subfigure}
\begin{subfigure}{0.9\textwidth}
    \centering
    \includegraphics[scale=0.56]{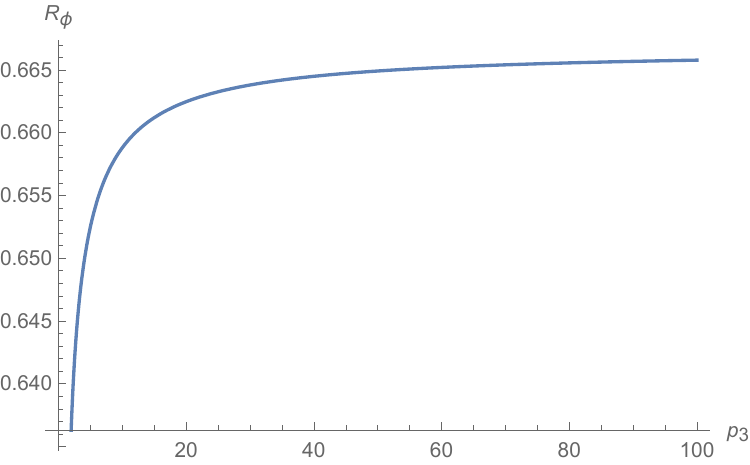}
    \caption{Plot of $R_\phi$ against $p_3$ for the gauging $(2,2,p_3)$ with one adjoint chiral.}
\end{subfigure}
\caption{Plots of $\epsilon_1$, $\epsilon_3$ and $R_\phi$ of $(2,2,p_3)$ theories glued by $\mathcal{N}=1$ gauging with one adjoint chiral multiplet. We can see that $\epsilon_1=\epsilon_2$, and $\epsilon_3$ lie in the range $\left(-\frac{1}{3},0\right)$. Furthermore, $R_\phi$ is always larger than $1/3$. Thus, we conclude that for all values of $p_3$, the gauged theory flows to an SCFT in the infrared with $a = c$.}
\label{fig:1adj_22p}
\end{figure}

\begin{figure}[H]
    \centering
\begin{subfigure}{0.45\textwidth}
    \includegraphics[scale=0.45]{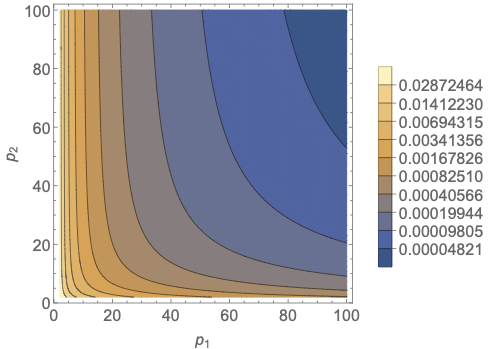}
    \caption{Contour plot of $-\epsilon_1$ on $(p_1,p_2)$ plane.}
\end{subfigure}
\begin{subfigure}{0.45\textwidth}
    \hfill\includegraphics[scale=0.45]{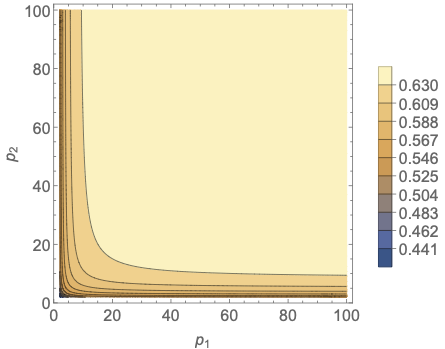}
    \caption{Contour plot of $R_\phi$ on $(p_1,p_2)$ plane.}
\end{subfigure}
\caption{Contour plots of $-\epsilon_1$ and $R_\phi$ of $\mathcal{D}_{p_{1}}(G)$ and $\mathcal{D}_{p_{2}}(G)$ theories glued by $\mathcal{N}=1$ gauging with one adjoint chiral multiplet. $\epsilon_1$ lies in the range  $\left(-\frac{1}{3},0\right)$, as does $\epsilon_2$, and $R_\phi$ is always larger than $1/3$; thus any such gauging satisfies the unitarity bound.}
\label{fig:1adj_2}
\end{figure}

Similarly to the discussion in Section \ref{sec:Neq1gluings}, analytically solving the coupled quadratic equations to obtain the mixing coefficients $\epsilon_i$ that maximize $a$ is technically challenging. Instead, we numerically explore the $p_i$-dependence of the solutions to the $a$-maximization condition. In Figure \ref{fig:1adj_2}, we numerically study the values of $\epsilon_1$, $\epsilon_2$, and $R_\phi$ of the gauged theories with $(p_1,p_2)$, for generic $p_1$ and $p_2$. We find that for all values of $p_1$ and $p_2$, the unitarity conditions on the operators are satisfied and hence the theory flows to an interacting SCFT with $a = c$ in the infrared. 
Similarly, Figure \ref{fig:1adj_22p} shows the numerical plots for $\epsilon_1=\epsilon_2$, $\epsilon_3$, and $R_\phi$ of the theories corresponding to $(2,2,p_3)$, with a generic $p_3$. Here, the asymptotic behavior in the large $p_3$ limit demonstrates that the unitarity bounds will be satisfied for all values of $p_3$. 
Finally, in Table \ref{tb:1adj_fin}, we write the $\epsilon_i$ and $R_\phi$ for the gaugings for the remaining cases in equation \eqref{eqn:1adjg}.
From this numerical analysis we successfully observe that all $\mathcal{N}=1$ gaugings with more than one $\mathcal{D}_p(G)$, such that $\gcd(p_i, h_G^\vee)=1$, and with an additional adjoint chiral multiplet flow to superconformal field theories with $a = c$ in the infrared.

We note that any set of $p_i$ that saturates the inequality in equation \eqref{eq:asymp_free_adj} has 
\begin{align}
    \epsilon_i=0,\quad R_\phi=\frac{2}{3}.
\end{align}
This reflects the fact that these theories have vanishing one-loop $\beta$-function and thus the theory does not flow. The $\mathcal{N}=2$ superpotential in equation \eqref{eq:N=2suppot} is marginal in these cases, and so the $\widehat{\Gamma}(G)$ theories, that are studied in \cite{Kang:2021lic}, and the theories with the same gauged $\mathcal{D}_p(G)$ and the same matter content, but without the superpotential term in equation \eqref{eq:N=2suppot}, are connected through a conformal manifold. This conformal manifold is parametrized by exactly marginal operators, which can be determined by enumerating marginal operators and symmetries \cite{Leigh:1995ep, Green:2010da}. 
The marginal operators include
\begin{align}
    \trace \phi^3 \ , \quad \trace \phi \mu_i \ .
\end{align}
A number of these marginal operators become marginally irrelevant after combining with the generators of the Abelian flavor symmetry, however we find that there is always at least one exactly marginal operator which in fact corresponds to the $\CN=2$ gauge coupling upon suitable normalization. For some choices of $\{p_i\}$, one can have additional marginal operators formed out of the products of Coulomb branch operators. 
We will revisit the study of marginal operators in \cite{OPSPEC,LANDSCAPE}.

\begin{table}[H]
\begin{threeparttable}
    \centering
    \renewcommand{\arraystretch}{1.2}
    $
    \begin{array}{c|ccccc}
    \toprule
        \{p_i\} & \epsilon_1 & \epsilon_2 & \epsilon_3 & \phantom{B}\epsilon_4\phantom{B} & R_\phi \\
    \midrule
        (1,2,3,3) & - & -0.0108332 & -0.00721553 & -0.00721553  & 0.626148 \\
        (1,2,3,4) & - & -0.00535965 & -0.00357149 & -0.00267712 & 0.645958 \\
        (1,2,3,5) & - & -0.00213347 & -0.00142206 & -0.00085290 & 0.658253 \\
        (1,2,3,6), (1, 2,4,4), (1, 3,3,3) & - & 0 & 0 & 0 & 2/3 \\
        (2,2,2,2) & 0 & 0 & 0 & 0 & 2/3
    \\\bottomrule
    \end{array}
    $
\end{threeparttable}
\caption{In this table, we write the values for the $\epsilon_i$ and the R-charge of the adjoint chiral multiplet, $R_\phi$, after gauging together the $\mathcal{D}_{p_i}(G)$ with one additional adjoint chiral multiplet. We can see that the values of these parameters, after doing the $a$-maximization to find the superconformal R-symmetry, are such that we obtain an interacting SCFT with $a = c$ in the infrared.}
    \label{tb:1adj_fin}
\end{table}

\subsection{Mass deformations of \texorpdfstring{$\mathcal{N}=2$}{N=2} \texorpdfstring{$\widehat{\Gamma}(G)$}{Gammahat(G)} theory}
\label{sec:massdef}

In this section, we explore the connection to the mass deformations of the 4d $\mathcal{N}=2$ SCFTs $\widehat{\Gamma}(G)$ considered in \cite{Kang:2021lic}. These $\widehat{\Gamma}(G)$ theories are associated to a pair of ADE groups, $\Gamma$ and $G$. When $\Gamma = D_4, E_6, E_7, E_8$ these theories correspond to an $\mathcal{N}=2$ diagonal gauging of the flavor symmetry of a collection of $\mathcal{D}_{p_i}(G)$ with respectively
\begin{align}
    (p_1, p_2, p_3, p_4) = (2,2,2,2), (1,3,3,3), (1,2,4,4), (1,2,3,6) \,.
\end{align}
These are precisely the same collections of $p_i$ as in equation \eqref{eqn:1adjg} that have vanishing one-loop $\beta$-function. When the gauging further satisfies $\gcd(p_i, h_G^\vee) = 1$ for all $p_i$ then the resulting $\mathcal{N}=2$ SCFTs have $a = c$. 

It is clear that the $\mathcal{N}=2$ gaugings and the $\mathcal{N}=1$ gaugings together with one additional adjoint chiral multiplet are distinct theories; the former has the additional superpotential term
\begin{equation}
    W=\sum_i\trace\mu_i\phi \,,
\end{equation}
from the $\mathcal{N}=2$ vector multiplet, and where $\phi$ is the adjoint-valued scalar field inside that vector multiplet. From the perspective of the $(\mathcal{N}=1)$-gauged theory, the terms in this superpotential are marginal operators that contribute to an $\mathcal{N}=1$ conformal manifold on which the $\widehat{\Gamma}(G)$ theory sits at a special point. Both theories have a relevant operator that gives a mass to either $\phi$ in the $\mathcal{N}=2$ vector multiplet, or the additional adjoint chiral $\phi$, respectively. It is natural to ask whether the two theories obtained via these two mass deformations, which integrate out the adjoint chirals, lead to the same infrared SCFTs.

First, we introduce a mass term for the adjoint $\mathcal{N}=1$ chiral multiplet $\phi$ inside of the $\mathcal{N}=2$ vector multiplet
\begin{equation}
    W = \trace\phi^2 \,.
\end{equation}
After adding such a mass term and flowing into the infrared we obtain an $\mathcal{N}=1$ theory where $\phi$ is integrated out. The $\mathcal{N}=1$ deformations of $\mathcal{N}=2$ theories through a mass term were studied in general in \cite{Tachikawa:2009tt}. The infrared $U(1)_R$ charge after the mass deformation is fixed to be
\begin{align}
    \begin{split}
        R_{\textrm{IR}}=\frac{1}{2}R_{\mathcal{N}=2}+I_3 \,,
    \end{split}
    \label{eq:m_def_R}
\end{align}
solely by the mass term in the superpotential. The central charges of the infrared $\mathcal{N}=1$ theory thus differ by a factor of $27/32$ from the central charges of the original $\mathcal{N}=2$ theory. 

On the other hand, the infrared R-symmetry, shown in equation \eqref{eq:IR_R}, of the $(\mathcal{N}=1)$-gauged theory satisfies the R-gauge-gauge anomaly-free condition
\begin{align}
    \begin{split}
        1+\sum_i\left(\frac{1}{3}-\epsilon_i\right)\left(-\frac{p_i-1}{p_i}\right)=0 \,.
    \end{split}
\end{align}
First, let us consider the cases where all of the $p_i$ are identical: where they are either $(2,2,2,2)$ or $(3,3,3)$. Each of the $\epsilon_i$ are then the same and the anomaly-free condition fixes that
\begin{equation}
    \epsilon_i = \left.\frac{1}{3} - \frac{p}{4(p-1)}\right|_{p=2} =-\frac{1}{6} \quad \text{or} \quad \epsilon_i = \left.\frac{1}{3} - \frac{p}{3(p-1)}\right|_{p=3} =-\frac{1}{6} \,,
\end{equation}
for the $\widehat{D}_4$ and $\widehat{E}_6$ cases, respectively. The operators $\trace  \mu_i \mu_j$ are exactly marginal and span a non-trivial conformal manifold for each of these theories.\footnote{There always exists at least one marginal operator $\trace \mu_i \mu_j$; chiral ring relations may reduce the dimension of the conformal manifold from the naive expectation.} Going to the point on the conformal manifold where the superpotential deformation is
\begin{align}
    \begin{split}
        \Delta W= \trace \left(\sum_i \mu_i\right)^2 \,,
    \end{split}
\end{align}
then we observe the mass-deformed $\widehat{D}_4(G)$ or $\widehat{E}_6(G)$ theories. 

In general, when not all of the $p_i$ are identical, the $\mathcal{N}=1$ gauging and the mass-deformation after $\mathcal{N}=2$ gauging yield different SCFTs with different central charges. 
For the $(\mathcal{N}=1)$-gauging when $(p_1, p_2, p_3) = (2, 4, 4)$ we find that
\begin{equation}
    \epsilon_1 = \frac{1}{3}(\sqrt{10} - 4), \qquad \epsilon_2 = \epsilon_3 = \frac{1}{9}(2 - \sqrt{10}),
\end{equation}
and thus 
\begin{equation}
    a = \frac{1}{144}(29 + 20\sqrt{10})\text{dim}(G).
\end{equation}
We can see that the operators 
\begin{align}
    \trace \mu_1 \mu_2,\quad\trace \mu_1 \mu_3
\end{align}
are relevant operators of this SCFT, and further there are no exactly marginal operators. The other quadratic trace of the moment maps, $\trace \mu_2\mu_3$, is an irrelevant operator.\footnote{The operators of the form $\trace \mu_i^2$ are projected out by the chiral ring relations of each individual $\mathcal{D}_p(G)$. Only the mixed operators $\trace \mu_i \mu_{j \neq i}$ survive, as we have discussed near equation \eqref{eq:trmuk0}.}
Let us now consider further deformations by these two relevant operators, sequentially. The resulting infrared SCFT has
\begin{equation}
    \epsilon_1 = \epsilon_2 = \epsilon_3 = -\frac{1}{6} \,,
\end{equation}
and the central charge $a$ is 
\begin{equation}
    a = \frac{81}{128}\text{dim}(G) = \frac{27}{32} a(\widehat{E}_7(G)) \,.
\end{equation}
We can see that this theory has the same central charge as the mass-deformation of the $\widehat{E}_7(G)$ theory. This is as we expect because there are three marginal operators $\trace \mu_i\mu_{j \neq i}$ and only one of them combines with the flavor current multiplet to become marginally irrelevant; in turn, we are left with two remaining \cite{Leigh:1995ep, Green:2010da}. These leftover marginal operators will span an at least two-dimensional conformal manifold that connects the two theories.

A similar analysis can be done for the  $(\mathcal{N}=1)$-gauging of the  $\mathcal{D}_{p_i}(G)$ theories with $(p_1, p_2, p_3) =(2,3,6)$ and we obtain irrational R-charge mixing coefficients
\begin{align}\label{eqn:e8mix}
    \begin{split}
        \epsilon_1\sim -0.284524\ ,\ \epsilon_2\sim-0.182519\ ,\ \epsilon_3\sim-0.0832703\ , 
    \end{split}
\end{align}
and corresponding central charge
\begin{align}
    \begin{split}
        a\sim 0.722376\ \textrm{dim}(G) \,.
    \end{split}
\end{align}
At the same time, the infrared limit of the mass-deformed $\widehat{E}_8(G)$ has its $a$ central charge by
\begin{align}
    \begin{split}
        a=\frac{27}{32}\,a({\widehat{E}_8(G)})
        = \frac{45}{64}\,\textrm{dim}(G)
        \sim 0.703125\ \textrm{dim}(G)
    \end{split}
\end{align}
which is smaller than that of the $(\mathcal{N}=1)$-gauged theory without superpotential. It agrees with the fact that each moment map operator $\mu_i$ has its R-charge given by \
\begin{align}
    \begin{split}
        R(\mu_i)=\frac{4}{3}+2\epsilon_i \,.
    \end{split}
\end{align}
We can see from the mixing coefficients in equation \eqref{eqn:e8mix} that $R(\mu_1)$ and $R(\mu_2)$ are both strictly less than one. We find that
\begin{align}
    \trace \mu_1 \mu_2,\quad \trace \mu_1 \mu_3 \,,
\end{align}
are relevant operators. 
We can turn on a superpotential deformation by either of these operators to trigger a renormalization group flow to an infrared fixed point. When we choose $\trace \mu_1 \mu_2$ the fixed point is the mass-deformed $\widehat{E}_8(G)$ theory; however, if we turns on $\trace \mu_1 \mu_3$, then the IR SCFT after the flow still has $\trace \mu_1 \mu_2$ as a relevant operator. Subsequently triggering a flow by this operator leads, again, to the mass-deformed $\widehat{E}_8(G)$ theory. 

This analysis explains why the $(\mathcal{N}=1)$-gauged theory has a central charge $a$ which is at least as large as the central charge of the mass-deformed $\widehat{\Gamma}(G)$ theory; it is required by the $a$-theorem \cite{Komargodski:2011vj}.

\subsection{\texorpdfstring{$\mathcal{N}=1$}{N=1} theory with two adjoint chirals: A Lagrangian model}
\label{sec:lagrangian}

Thus far all the $\mathcal{N}=1$ SCFTs with $a = c$ that we obtain are rather exotic and non-Lagrangian; all of them involve the Argyres--Douglas theories $\mathcal{D}_p(G)$. In fact, there is a simple Lagrangian gauge theory with $a = c$: consider a gauge theory with gauge group $G$ and two adjoint chiral multiplets. We can consider this theory as a special case where we gauge \emph{zero} $\mathcal{D}_p(G)$ theories, together with two adjoint chiral multiplets $\phi_1$ and $\phi_2$. We can see that this configuration is asymptotically-free from equation \eqref{eq:asymp_free_adj}. From the anomaly-free condition for the R-symmetry, we get
\begin{align}
    \mathrm{Tr} R GG = 0  \quad \Longleftrightarrow \quad h_G^{\vee}\left(1 + 2(R_\phi - 1)\right) = 0 \,, 
\end{align}
where we use the symmetry to write $R_{\phi_1} = R_{\phi_2}=R_\phi$. The anomaly cancellation enforces that the R-charges for the two adjoint chiral multiplets are 
\begin{align}
    R_{\phi_1} = R_{\phi_2} = \frac{1}{2} \,. 
\end{align}
Therefore, we find
\begin{align}
    16(a - c) = \mathrm{Tr} R = \left( 1 + 2 \left(\frac{1}{2}-1\right) \right) \mathrm{dim}(G) = 0 \ , 
\end{align}
and thus the theory realizes $a=c$. The theory has an $SU(2)$ flavor symmetry rotating the two adjoint chiral multiplets, and the central charges are given by
\begin{align}
    a = c =\frac{9}{32} \left( 1^3 + 2 \left(\frac{1}{2}-1\right)^3 \right) \mathrm{dim}(G) = \frac{27}{128} \mathrm{dim}(G) \,. 
\end{align}
This theory also belongs to the conformal manifold of the theory obtained starting from $\mathcal{N}=4$ super  Yang--Mills with gauge group $G$ and triggering an RG-flow by adding a mass term for one of the three adjoint chiral multiplets inside the $\mathcal{N}=4$ vector multiplet. As expected from equation \eqref{eqn:coolratio}, the central charges of the infrared $\mathcal{N}=1$ theory are $27/32$ times the central charges of the $\mathcal{N}=4$ theory \cite{Tachikawa:2009tt}.

\subsection{\texorpdfstring{$\mathcal{N}=1$}{N=1} gluing with two adjoint chirals}
\label{sec:2adjoints}

The sets of $\mathcal{D}_p(G)$ theories that can be gauged together when we include two adjoint chiral multiplets on the gauge node are highly restricted, as we can see from equation \eqref{eq:asymp_free_adj}. In the simplest case, we consider a single $\mathcal{D}_p(G)$ theory for any choice of $p$. Then we obtain the asymptotically-free theory given by
\vspace{-8mm}
\begin{align}
\begin{aligned}
    \begin{tikzpicture}
      \node[gaugeN1] (s0) {$G$};
      \node[d2] (c2) [left=0.6cm of s0] {$\mathcal{D}_{p}(G)$};
      \draw (s0.west) -- (c2.east);
       \draw[dashed, ->] (s0) to[out=40, in=320, looseness=4] (s0);
      \draw[dashed, ->] (s0) to[out=56, in=304, looseness=7] (s0);
    \end{tikzpicture}
\end{aligned} \,.
\end{align}
\vspace{-14mm}

\noindent A single $\mathcal{D}_p(G)$ theory gauged with two adjoint chiral multiplets attached has its infrared R-charge given by
\begin{align}
    \begin{split}
        R=R_0+\epsilon \mathcal{F} \,,
    \end{split}
\end{align}
where $\epsilon$ and the R-charge of adjoint chiral multiplets $\phi_1$ and $\phi_2$ are 
\begin{align}
    \begin{split}
        \epsilon&=\frac{-8p^3-2p^2+p+1+2p\sqrt{16p^4+8p^3-11p^2+3}}{3(8p^3-7p^2-2p+1)}\ , \\
        R_{\phi_1} = R_{\phi_2} &=\frac{20p^2-p-3-\sqrt{16p^4+8p^3-11p^2+3}}{3(8p^2+p-1)} \,.
    \end{split}
\end{align}
It is straightforward to check that the operators satisfy the unitarity conditions for any value of $p$, and thus each theory flows in the infrared to an interacting SCFT with $a = c$, if $\gcd(p, h_G^\vee) = 1$. The resulting SCFTs have a host of relevant operators 
\begin{align}\label{eqn:opop}
    \trace \phi_1^2\,,\,\, 
    \trace \phi_1\phi_2 \,,\,\,
    \trace \phi_2^2 \,,\,\, 
    \trace \phi_1^3 \,,\,\, 
    \trace \phi_1^2\phi_2 \,,\,\, 
    \trace \phi_1\phi_2^2 \,,\,\, 
    \trace \phi_2^3 \,,\,\,
    \trace \mu \phi_1 \,,\,\, 
    \trace \mu \phi_2 \,.
\end{align}
Each of these operators provides a superpotential deformation that triggers a renormalization group flow to a new infrared SCFT. 
Let us note that when $G=SU(2)$, the cubic operators in equation \eqref{eqn:opop} are absent. 
The landscape charted by superpotential deformations involving these relevant operators is one of the subjects of study in \cite{LANDSCAPE}.

The only other possibility for gauging together $\mathcal{D}_p(G)$ with two adjoint chiral multiplets is a theory with vanishing one-loop $\beta$-function that is obtained by gauging two $\mathcal{D}_2(G)$ theories. The resulting theory is of the form
\begin{align}\label{eqn:2adj22}
\begin{aligned}
    \begin{tikzpicture}
      \node[gaugeN1] (s0) {$G$};
      \node[d2] (c2) [left=0.6cm of s0] {$\mathcal{D}_{2}(G)$};
      \node[d2] (c3) [right=0.6cm of s0] {$\mathcal{D}_{2}(G)$};
      \draw (s0.west) -- (c2.east);
      \draw (s0.east) -- (c3.west);
       \draw[dashed, ->] (s0) to[out=130, in=410, looseness=4] (s0);
      \draw[dashed, ->] (s0) to[out=146, in=394, looseness=7] (s0);
    \end{tikzpicture} \,.
\end{aligned}
\end{align}
As it is discussed in Section \ref{sec:Neq1gluings}, the gaugings that saturate the inequality in equation \eqref{eq:asymp_free} do not necessarily lead to an interacting superconformal field theory, as they may not have any exactly marginal operators.\footnote{While such conformal gaugings may not necessarily admit any exactly marginal operators, we see in \cite{OPSPEC} that each gauging appearing in Table \ref{tbl:confgaugep} does.} 
However, when two $\mathcal{D}_2(G)$ theories are glued together with two adjoint chiral multiplets, we expect a non-trivial SCFT as there are now marginal operators built out of the adjoint chiral multiplets. There are eight marginal operators
\begin{align}
    \operatorname{Tr}\mu_i \phi_1 \,, \quad \operatorname{Tr}\mu_i \phi_2 \,, \quad \operatorname{Tr}\phi_1^3 \,, \quad \operatorname{Tr}\phi_1^2 \phi_2 \,, \quad \operatorname{Tr}\phi_1 \phi_2^2 \,, \quad \operatorname{Tr}\phi_2^3 \,.
\end{align}
For group-theoretic reasons, a number of these operators may not exist, for example, for $G = SU(2)$, the four operators that are cubic in $\phi_i$ are not present due to the absence of a cubic Casimir. Since $\gcd(2, h_G^\vee) = 1$ is required to obtain a theory with identical central charges $a=c$, it is necessary to have $G = SU(2n+1)$ to ensure $a=c$. Thus, the cubic marginal operators are present in the theories with $a=c$.
Among the eight marginal operators, at most five of them may become marginally irrelevant as they combine with the generators of the $SU(2)\times U(1)^2$ flavor symmetry. The remaining operators are exactly marginal and contribute to the conformal manifold with dimension at least three.\footnote{The marginal operators belonging to the $\mathcal{D}_p(G)$ theory before gauging will also contribute to the conformal manifold.}

\section{Beyond \texorpdfstring{\boldmath{$a = c$}}{a=c}: Including conformal matter}\label{sec:beyondac}

In the current paper, we mainly focused on $\CN=1$ SCFTs with $a=c$, but let us discuss a natural generalization of our setup that may have $a \neq c$.
In \cite{Kang:2021lic}, the authors discussed 4d $\mathcal{N}=2$ SCFTs called $\widehat{\Gamma}(G)$, for all choices of ADE Lie algebra $\Gamma$. These theories were constructed by taking a collection of Argyres--Douglas $\mathcal{D}_{p_i}(G)$ theories and a collection of $(G,G)$ conformal matter theories\footnote{These theories are sometimes known as the generalized $(G\times G)$ bifundamental. They are obtained by compactifying the relevant six-dimensional minimal conformal matter theory \cite{DelZotto:2014hpa,Ohmori:2015pua,Baume:2021qho} on a torus. Some $\mathcal{N}=1$ gaugings of $\mathcal{N}=2$ conformal matter have been explored in \cite{Apruzzi:2018oge}.} and performing $\mathcal{N}=2$ conformal gaugings of all of the $G$ flavor symmetries.\footnote{The resulting theories can have residual  flavor symmetries if the $\mathcal{D}_{p_i}(G)$ or conformal matter theories have a larger flavor symmetry than $G$ or $(G\times G)$, respectively.} In \cite{Kang:2021lic}, it was found that the classification of all possible such gaugings is equivalent to the ADE classification problem, and thus the solutions are labeled by an ADE algebra $\Gamma$. For $\Gamma = D_4, E_6, E_7, E_8$, this gauging involved the introduction of a single gauge node, and no copies of the conformal matter theory, however for $\Gamma = A_{n-1 \geq 0}$ and $\Gamma = D_{n \geq 5}$ the gauging involved adding either of ring of $n$ gauge nodes, or a linear chain of $(n-3)$ gauge nodes, respectively.

In this paper, we have focused on gauging together a collection of $\mathcal{D}_{p_i}(G)$, possibly with additional adjoint chiral multiplets, via a \emph{single} $\mathcal{N}=1$ vector multiplet. We study this particular setup because we desire to explore 4d $\mathcal{N}=1$ SCFTs with identical central charges, $a = c$. In the setup of $\widehat{\Gamma}(G)$ theories, it is only in the cases where $\Gamma = D_4, E_6, E_7, E_8$, and with specific values of $G$, that we obtain 4d $\mathcal{N}=2$ SCFTs with $a = c$. In a similar vein, we would expect that any $\mathcal{N}=1$ gauging involving multiple gauge nodes would create a deviation from $a = c$ due to the included conformal matter theories. However, if we are willing to chart the landscape beyond $a = c$, there is no a priori problem with considering such a class of $\mathcal{N}=1$ SCFTs, obtained from $\mathcal{N}=1$ gaugings of a collection of $\mathcal{D}_{p_i}(G)$ theories and a collection of $(G\times G)$ conformal matter theories. 

A natural starting point for studying $\mathcal{N}=1$ gaugings involving both Argyres--Douglas and conformal matter theories could be to consider mass-deformations associated to the adjoint chiral multiplets in the $\mathcal{N}=2$ theories $\widehat{A}_{n-1\geq 1}(G)$ and $\widehat{D}_{n \geq 5}(G)$. This would give a interesting collection of theories with IR central charge being $27/32$ of the UV central charges \cite{Tachikawa:2009tt}. Here we also would like to remind the reader that a mass-deformation of chiral adjoint in a vector multiplet of $\CN=2$ theory is not equivalent to the $\CN=1$ gauging due to the superpotential generated by the mass term. We can consider various configurations corresponding to different choices of $\CN=1$ and $\CN=2$ gaugings for the multiple gauge nodes as in the case of $\CN=1$ class $\mathcal{S}$ theories \cite{Bah:2012dg, Benini:2009mz}. 

The number of options for $\CN=1$ gaugings grows rapidly with the number of conformal matter theories involved. We introduce some simplifying notation: define $\mathcal{C}_n$ as all possible sets of $p_i$ that solve the inequality
\begin{equation}\label{eqn:cmineq}
    \mathcal{C}_n\, :\quad \sum_i \frac{p_i - 1}{p_i} \leq 3 - n \,.
\end{equation}
This is just the condition in equation \eqref{eq:asymp_free_adj}, which dictates which sets of $\mathcal{D}_{p_i}(G)$ can be ($\mathcal{N}=1$)-gauged together with $n$ additional adjoint chiral multiplets. The sets of $p_i$ that belong to $\mathcal{C}_0$ are listed in Tables \ref{tbl:asympfreep} and \ref{tbl:confgaugep}. For $\mathcal{C}_1$ and $\mathcal{C}_2$, which will be of relevance in this section, we summarize which sets of $p_i$ belong to them in Table \ref{tbl:C1C2}. We write $\mathcal{C}_n$ in a quiver, connected to a gauge node, as a shorthand for all of the quivers involving all sets $\{p_i\}$ inside $\mathcal{C}_n$. For example, the quiver
\begin{equation}
    \begin{aligned}
        \begin{tikzpicture}
            \node[d2] (s0) {$\mathcal{C}_2$};
            \node[gaugeN1] (s1) [right=0.5cm of s0] {$G$};
            \draw (s0.east) -- (s1.west);
        \end{tikzpicture} \,,
    \end{aligned}
\end{equation}
is used to succinctly refer to the two quivers 
\begin{equation}
    \begin{aligned}
        \begin{tikzpicture}
            \node[d2] (s0) {$\mathcal{D}_p(G)$};
            \node[gaugeN1] (s1) [right=0.5cm of s0] {$G$};
            \draw (s0.east) -- (s1.west);
        \end{tikzpicture} \,, \qquad\qquad 
        \begin{tikzpicture}
            \node[d2] (s0) {$\mathcal{D}_2(G)$};
            \node[d2] (s2)  [right=0.5cm of s1] {$\mathcal{D}_2(G)$};
            \node[gaugeN1] (s1) [right=0.5cm of s0] {$G$};
            \draw (s0.east) -- (s1.west);
            \draw (s1.east) -- (s2.west);
        \end{tikzpicture} \,.
    \end{aligned}
\end{equation}

\begin{table}[H]
    \centering
    \begin{threeparttable}
    \renewcommand{\arraystretch}{1.3}
    \begin{tabular}{ccc}
        \toprule
        \multirow{2}{*}{\phantom{N}Name\phantom{N}} & \multicolumn{2}{c}{$\{p_i\}$} \\\cline{2-3}
        & \phantom{Nam}Asymptotically-free gauging\phantom{Nam} & \phantom{Nam}Conformal gauging\phantom{Nam} \\\midrule
        $\mathcal{C}_1$ & $\begin{gathered} \{p \geq 2\} \quad \{p_1 \geq 2, p_2 \geq 2\} \\ \{2, 2, p \geq 2\} \quad \{2,3,4\} \quad \{2,3,5\} \end{gathered}$ & $\begin{gathered} \{2,2,2,2\} \quad \{3, 3, 3\} \\ \{2,4,4\} \quad \{2,3,6\} \end{gathered}$  \\\midrule
        $\mathcal{C}_2$ & $\{p \geq 2\}$ & $\{2, 2\}$ \\\bottomrule
    \end{tabular}
    \end{threeparttable}
    \caption{We define $\mathcal{C}_n$ as the set of $\{p_i\}$ satisfying the inequality in equation \eqref{eqn:cmineq}. Here, we list all the elements inside $\mathcal{C}_1$ and $\mathcal{C}_2$. These are identical to the collections of $p_i$ found for gauging together with, respectively, one and two adjoint chiral multiplets in Section \ref{sec:withchirals}. The conformal gaugings saturate the inequality while the asymptotically-free gaugings do not.}
    \label{tbl:C1C2}
\end{table}

Suppose that we wish to take the diagonal gauging of $n$ flavor symmetries $G$ coming from $(G\times G)$ conformal matter, together with the $G$ of a collection of $\mathcal{D}_{p_i}(G)$. The constraint for a gauging which is either asymptotically free or conformal is\footnote{It is actually possible to obtain an interacting theory in the IR even when the gauge coupling is not asymptotically free when the gauge group is given as a product form \cite{Barnes:2005zn}. This is due to the mixing of a non-anomalous $U(1)$ symmetry with the R-symmetry in the infrared. This can result in some of the gauge couplings being ``dangerously irrelevant''.}
\begin{equation}\label{eqn:DpCM}
    \sum_{i} \frac{2(p_i - 1)}{p}h_G^\vee + 2nh_G^\vee\, \leq\, 6h_G^\vee \,,
\end{equation}
where we have used that the flavor central charge of each $G$ flavor symmetry of conformal matter is $2h_G^\vee$. We can see that this is equivalent to the inequality in equation \eqref{eqn:cmineq}, and thus a gauge node obtained by the $\mathcal{N}=1$ diagonal gauging involving $n$ conformal matter factors can also involve any collection of $\mathcal{D}_{p_i}(G)$ from $\mathcal{C}_n$. \\

\begin{figure}[H]
    \centering
    \begin{subfigure}[b]{0.3\textwidth}
        \centering
        \begin{tikzpicture}
            \node[d2] (s0) {$\mathcal{C}_1$};
            \node[gaugeN1] (s1) [right=0.5cm of s0] {$G$};
            \node[gaugeN1] (s2) [right=0.5cm of s1] {$G$};
            \node[d2] (s3) [right=0.5cm of s2] {$\mathcal{C}_1$};
            \draw (s0.east) -- (s1.west);
            \draw (s1.east) -- (s2.west);
            \draw (s2.east) -- (s3.west);
        \end{tikzpicture}
    \end{subfigure}
    \begin{subfigure}[b]{0.3\textwidth}
        \centering
        \begin{tikzpicture}
            \node[d2] (s0) {$\mathcal{C}_2$};
            \node[gaugeN1] (s1) [right=0.5cm of s0] {$G$};
            \node[gaugeN1] (s2) [right=0.5cm of s1] {$G$};
            \node[d2] (s3) [right=0.5cm of s2] {$\mathcal{C}_2$};
            \draw (s0.east) -- (s1.west);
            \draw[double, double distance=1mm] (s1.east) -- (s2.west);
            \draw (s2.east) -- (s3.west);
        \end{tikzpicture}
    \end{subfigure}
    \begin{subfigure}[b]{0.3\textwidth}
        \centering
        \begin{tikzpicture}
            \node[gaugeN1] (s1) {$G$};
            \node[gaugeN1] (s2) [right=0.5cm of s1] {$G$};
            \draw[double, double distance=2mm] (s1.east) -- (s2.west);
            \draw (s1.east) -- (s2.west);
        \end{tikzpicture}
    \end{subfigure} \vspace{0.5cm}\newline
    \begin{subfigure}[b]{0.40\textwidth}
        \centering
        \begin{tikzpicture}
            \node[gaugeN1] (s1) {$G$};
            \node[gaugeN1] (s2) [right=0.5cm of s1] {$G$};
            \node[d2] (s3) [right=0.5cm of s2] {$\mathcal{C}_1$};
            \draw (s1.east) -- (s2.west);
            \draw (s2.east) -- (s3.west);
            \draw (s1) to[out=140, in=220, looseness=4] (s1);
        \end{tikzpicture}
    \end{subfigure}
    \begin{subfigure}[b]{0.40\textwidth}
        \centering
        \begin{tikzpicture}
            \node[gaugeN1] (s1) {$G$};
            \node[gaugeN1] (s2) [right=0.5cm of s1] {$G$};
            \draw (s1.east) -- (s2.west);
            \draw (s1) to[out=140, in=220, looseness=4] (s1);
            \draw (s2) to[out=40, in=320, looseness=4] (s2);
        \end{tikzpicture}
    \end{subfigure}
    \caption{All possible asymptotically-free or conformal gaugings of an arbitrary collection of $\mathcal{D}_{p_i}(G)$ and $(G\times G)$ theories with two gauge nodes. A node with a solid border denotes an $\mathcal{N}=1$ gauge node, and a solid line between gauge nodes refers to the presence of $(G\times G)$ conformal matter. A line connecting a gauge node and $\mathcal{C}_n$ is a shorthand notation for all options of gaugings with the collections of $\mathcal{D}_{p_i}(G)$ theories as described in Table \ref{tbl:C1C2}.}
    \label{fig:twogauge}
\end{figure}
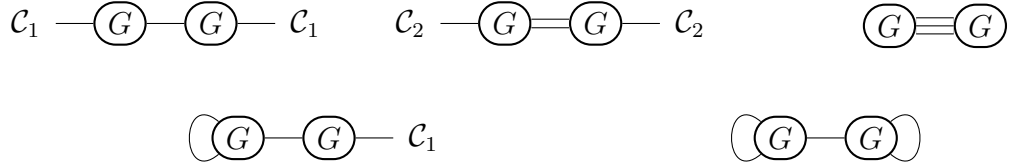

\begin{figure}[H]
    \centering
    \begin{subfigure}[b]{0.36\textwidth}
        \centering
        \begin{tikzpicture}
            \node[d2] (s0) {$\mathcal{C}_1$};
            \node[gaugeN1] (s1) [right=0.5cm of s0] {$G$};
            \node[gaugeN1] (s2) [right=0.5cm of s1] {$G$};
            \node[gaugeN1] (s3) [right=0.5cm of s2] {$G$};
            \node[d2] (s4) [right=0.5cm of s3] {$\mathcal{C}_1$};
            \node[d2] (s5) [above=0.5cm of s2] {$\mathcal{C}_2$};
            \draw (s0.east) -- (s1.west);
            \draw (s1.east) -- (s2.west);
            \draw (s2.east) -- (s3.west);
            \draw (s3.east) -- (s4.west);
            \draw (s2.north) -- (s5.south);
            \path (s1) to[out=140, in=220, looseness=4] (s1);
        \end{tikzpicture}
    \end{subfigure}
    \begin{subfigure}[b]{0.31\textwidth}
        \centering
        \begin{tikzpicture}
            \node[gaugeN1] (s1) {$G$};
            \node[gaugeN1] (s2) [right=0.5cm of s1] {$G$};
            \node[gaugeN1] (s3) [right=0.5cm of s2] {$G$};
            \node[d2] (s4) [right=0.5cm of s3] {$\mathcal{C}_1$};
            \node[d2] (s5) [above=0.5cm of s2] {$\mathcal{C}_2$};
            \draw (s1.east) -- (s2.west);
            \draw (s2.east) -- (s3.west);
            \draw (s3.east) -- (s4.west);
            \draw (s2.north) -- (s5.south);
            \draw (s1) to[out=140, in=220, looseness=4] (s1);
        \end{tikzpicture}
    \end{subfigure} 
    \begin{subfigure}[b]{0.30\textwidth}
        \centering
        \begin{tikzpicture}
            \node[gaugeN1] (s1) {$G$};
            \node[gaugeN1] (s2) [right=0.5cm of s1] {$G$};
            \node[gaugeN1] (s3) [right=0.5cm of s2] {$G$};
            \node[d2] (s5) [above=0.5cm of s2] {$\mathcal{C}_2$};
            \draw (s1.east) -- (s2.west);
            \draw (s2.east) -- (s3.west);
            \draw (s2.north) -- (s5.south);
            \draw (s1) to[out=140, in=220, looseness=4] (s1);
            \draw (s3) to[out=40, in=320, looseness=4] (s3);
        \end{tikzpicture}
    \end{subfigure} \vspace{0.5cm}\newline
    \begin{subfigure}[b]{0.36\textwidth}
        \centering
        \begin{tikzpicture}
            \node[d2] (s0) {$\mathcal{C}_2$};
            \node[gaugeN1] (s1) [right=0.5cm of s0] {$G$};
            \node[gaugeN1] (s2) [right=0.5cm of s1] {$G$};
            \node[gaugeN1] (s3) [right=0.5cm of s2] {$G$};
            \node[d2] (s4) [right=0.5cm of s3] {$\mathcal{C}_1$};
            \draw (s0.east) -- (s1.west);
            \draw[double, double distance=1mm] (s1.east) -- (s2.west);
            \draw (s2.east) -- (s3.west);
            \draw (s3.east) -- (s4.west);
            \path (s1) to[out=140, in=220, looseness=4] (s1);
        \end{tikzpicture}
    \end{subfigure}
    \begin{subfigure}[b]{0.36\textwidth}
        \centering
        \begin{tikzpicture}
            \node[d2] (s0) {$\mathcal{C}_2$};
            \node[gaugeN1] (s1) [right=0.5cm of s0] {$G$};
            \node[gaugeN1] (s2) [right=0.5cm of s1] {$G$};
            \node[gaugeN1] (s3) [right=0.5cm of s2] {$G$};
            \draw (s0.east) -- (s1.west);
            \draw[double, double distance=1mm] (s1.east) -- (s2.west);
            \draw (s2.east) -- (s3.west);
            \path (s1) to[out=140, in=220, looseness=4] (s1);
            \draw (s3) to[out=40, in=320, looseness=4] (s3);
        \end{tikzpicture}
    \end{subfigure} \vspace{0.5cm}\newline
    \begin{subfigure}[b]{0.36\textwidth}
        \centering
        \begin{tikzpicture}
            \node[gaugeN1] (s1) {$G$};
            \node[gaugeN1] (s2) [right=0.8cm of s1] {$G$};
            \node[gaugeN1] (s3) [above right=0.88cm of s1] {$G$};
            \node[d2] (s4) [right=0.5cm of s2] {$\mathcal{C}_2$};
            \node[d2] (s5) [left=0.5cm of s1] {$\mathcal{C}_2$};
            \node[d2] (s6) [above=0.5cm of s3] {$\mathcal{C}_2$};
            \draw (s1.east) -- (s2.west);
            \draw (s1.45) -- (s3.250);
            \draw (s2.135) -- (s3.290);
            \draw (s1.west) -- (s5.east);
            \draw (s2.east) -- (s4.west);
            \draw (s3.north) -- (s6.south);
            \path (s1) to[out=140, in=220, looseness=4] (s1);
        \end{tikzpicture}
    \end{subfigure}
    \begin{subfigure}[b]{0.36\textwidth}
        \centering
        \begin{tikzpicture}
            \node[gaugeN1] (s1) {$G$};
            \node[gaugeN1] (s2) [right=0.8cm of s1] {$G$};
            \node[gaugeN1] (s3) [above right=0.88cm of s1] {$G$};
            \node[d2] (s6) [above=0.5cm of s3] {$\mathcal{C}_2$};
            \draw[double, double distance=1mm] (s1.east) -- (s2.west);
            \draw (s1.45) -- (s3.250);
            \draw (s2.135) -- (s3.290);
            \draw (s3.north) -- (s6.south);
            \path (s1) to[out=140, in=220, looseness=4] (s1);
        \end{tikzpicture}
    \end{subfigure}
    \caption{All possible gaugings of conformal matter and $\mathcal{D}_p(G)$ with three gauge nodes. See the caption of Figure \ref{fig:twogauge} for an explanation of the notation.}
    \label{fig:threegauge}
\end{figure}
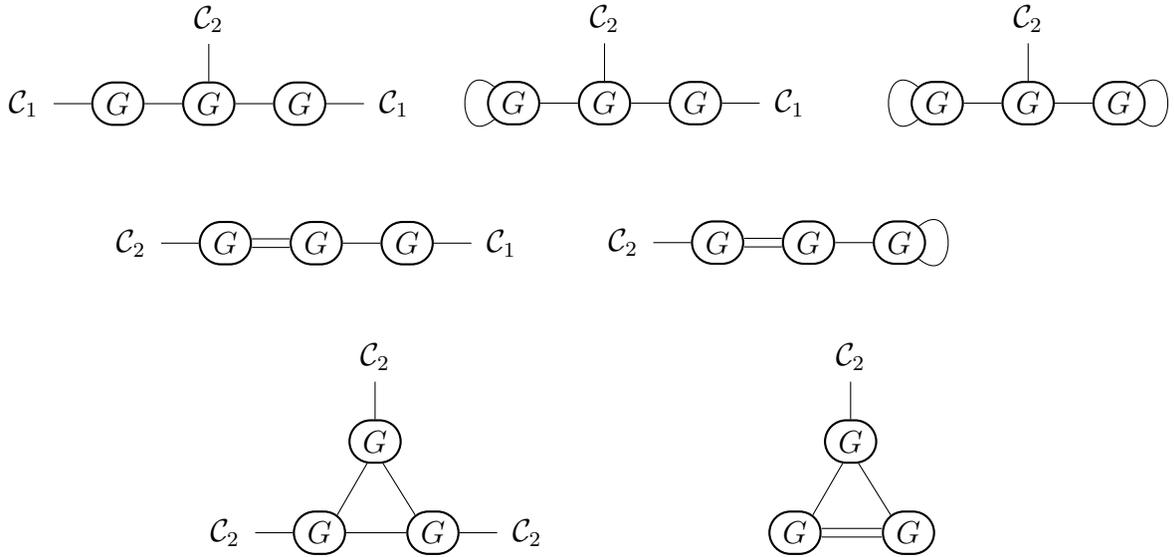

The solution of the classification problem for the $\mathcal{N}=1$ gaugings of $\mathcal{D}_{p_i}(G)$ and $(G\times G)$ conformal matter, which do not in general have identical central charges $a=c$, is left for future work. In fact, in addition to solving the classification problem for the gaugings, it is necessary to determine whether each quiver formed by the gauging flows in the infrared to an interacting SCFT. 
Here, let us discuss some preliminary examples including $\mathcal{D}_p(G)$ and $(G\times G)$ conformal matter theories as in Figures \ref{fig:twogauge} and \ref{fig:threegauge}.
We depict all such solutions involving precisely two gauge nodes in Figure \ref{fig:twogauge}. Similarly, we show all those configurations involving three gauge nodes in Figure \ref{fig:threegauge}. As depicted in these figures, we can see that there are in total five \emph{types} of diagrams involving two gauge nodes and seven types involving three gauge nodes, where each type may contain infinitely-many quivers once all the options of $\mathcal{C}_n$ are taken into account. In fact, the number of types grows rapidly with the number of gauge nodes; for example, for the case of four gauge nodes, there are already twenty-two types.

The gaugings which we have briefly introduced in this section form a broad class of $\mathcal{N}=1$ theories, typically without any non-Abelian flavor symmetry, although we do not expect them to have identical central charges. 
We leave a detailed analysis of this intriguing class of 4d $\mathcal{N}=1$ theories for future work.

\section{Discussion}
\label{sec:discussion}

Throughout the paper, we explored under what circumstances a collection of 4d $\mathcal{N}=2$ Argyres--Douglas $\mathcal{D}_{p_i}(G)$ theories can be diagonally gauged together via an $\mathcal{N}=1$ $G$-vector multiplet in such a way that the $\mathcal{N}=1$ theory in the infrared is an interacting SCFT with identical central charges, $a=c$. We also determined that the addition of one or two chiral multiplets transforming in the adjoint representation of $G$ is compatible with the $a = c$ property. Surprisingly, we discover that almost all asymptotically-free gaugings (and also almost all conformal gaugings when additional adjoint chirals are included) flow to IR SCFTs with $a = c$, when $G$ satisfies $\gcd(p_i, h_G^\vee) = 1$ for all $p_i$. This opens up a vast but well-curated landscape of minimally supersymmetric CFTs with exactly equal central charges. In this section, we discuss possible future directions. 

\begin{figure}[H]
    \centering
    \begin{tikzcd}[column sep=0.8in, row sep=0.8in, scale=1.2]
        & & \widehat{O}\arrow[rightarrow]{lld}[swap]{\Delta W\sim\trace\mu \phi_2\phantom{m}} \arrow[rightarrow]{ld}[description,pos=0.6]{\Delta W\sim \trace\phi_2^3} \arrow[rightarrow]{d}[description,pos=0.65]{\Delta  W\sim \trace\phi_1 \phi_2^2} \arrow[rightarrow]{rd}{\Delta  W\sim \trace\phi_2^2+M \trace\phi_1^2} & & \\
        ? & \widehat{E} & \widehat{D}\arrow[rightarrow]{d}[description]{\Delta  W\sim \trace\phi_1^{k-1}} & \widehat{A} \arrow[rightarrow]{d}[description]{\Delta  W\sim\trace\mu \phi_1 \phantom{m}} \arrow[rightarrow]{rd}{\Delta  W\sim\trace \phi_1^{k+1}} & \\
        & & D_k\, (k=3,4,5) & ? & A_k\, (2\leq k\leq 6)
    \end{tikzcd}
    \caption{Deforming the theories with one $\mathcal{D}_p(G)$ and two adjoint chiral multiplets $(\phi_1, \phi_2)$, that we refer to as $\widehat{O}$, via ADE-type superpotential. When the $\widehat{O}$ is deformed by $\trace\phi_2^2$, the $\trace\phi_1^2$ becomes free during the RG-flow and must be flipped by the flip-operator $M$. The question marks represent non-ADE deformations.}
    \label{fig:flows}
\end{figure}
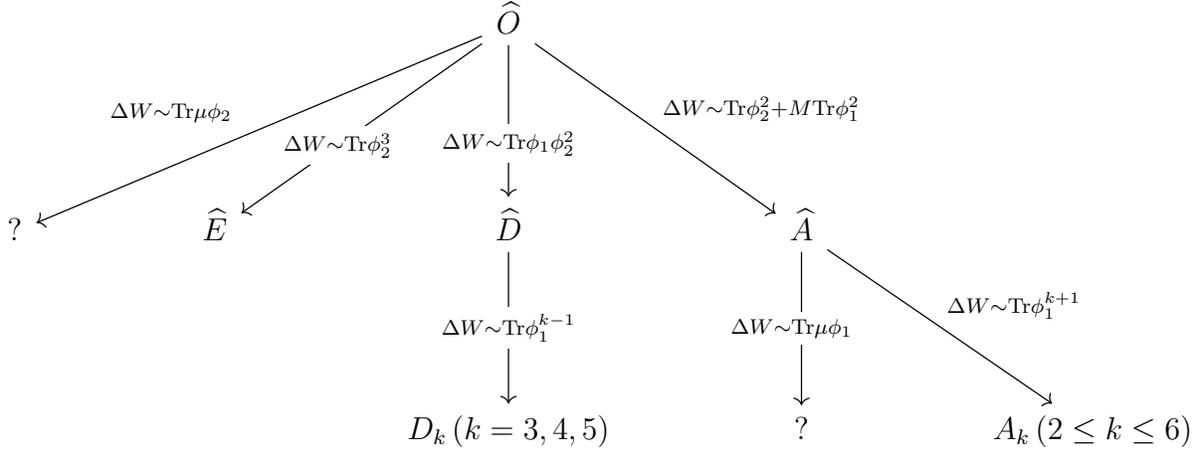

Beyond the constructions worked out in the current paper, we noted in Section \ref{sec:acgauging} that the addition of a superpotential deformation to the $(\mathcal{N}=1)$-gauging does not spoil the argument that the infrared SCFTs have $a = c$, as long as the assumption that there are no emergent Abelian flavor symmetries is not violated. This opens up an even vaster landscape of 4d $\mathcal{N}=1$ SCFTs with $a = c$. Particularly interesting examples in this class of theories are those obtained by deformation of a single $\mathcal{D}_p(G)$ theory gauged together with two adjoint chiral multiplets. In \cite{Intriligator:2003mi}, the authors showed that the superpotential deformations of SQCD with adjoint chirals has a classification that coincides with the ADE classification. Similar ADE-type deformations exist for $\CN=1$ theories constructed out of gauging $\mathcal{D}_p(G)$ with two adjoint chiral multiplets, as can be seen in Figure \ref{fig:flows}, and we explore these and other superpotential deformations of the $(\mathcal{N}=1)$-gaugings in \cite{LANDSCAPE}.

We have considered the 4d $\mathcal{N}=1$ SCFTs that live as the infrared fixed points of the diagonal $(\mathcal{N}=1)$-gauging of $G$ of a collection of $\mathcal{N}=2$ $\mathcal{D}_p(G)$ theories, when $G$ is an ADE Lie group. Two natural extensions of the current work present themselves immediately. First, the definition of the $\mathcal{D}_p(G)$ theories can be extended to include $G$ being of BCFG-type \cite{Carta:2021whq,Wang:2018gvb}. Second, the $\mathcal{D}_p(G)$ are the special case of the $\mathcal{D}_p^b(G)$ theories, where $b = h_G^\vee$ \cite{Wang:2015mra}. In this broader context, we can ask which collections of $\mathcal{D}_{p_i}^b(G \in \text{ABCDEFG})$ can be gauged together by an $\mathcal{N}=1$ vector multiplet in such a way that one obtains an interacting infrared SCFT with $a = c$ as was partially investigated in the $\CN=2$ setup in \cite{Closset:2020afy}. We leave this generalization for future work.

In another direction, some of the $\mathcal{N}=2$ SCFTs $\widehat{\Gamma}(G)$ have a fascinating connection to $\mathcal{N}=4$ super  Yang--Mills. It is discovered in \cite{Kang:2021lic, Buican:2020moo} that the Schur indices of particular $\widehat{\Gamma}(G)$ theories with $a = c$ can be written as the Schur index of $\mathcal{N}=4$ super  Yang--Mills with re-scaled fugacities.\footnote{It is found that the $\widehat{\Gamma}(G)$ theories without flavor symmetries, which is a superset of $\widehat{\Gamma}(G)$ theories with $a=c$, will all have their Schur indices to be the same with that of $\mathcal{N}=4$ SYM upto recaling fugacities.} Is there a sense in which the $\mathcal{N}=1$ SCFTs with $a = c$ that we obtain in this paper are related to $\mathcal{N}=4$ super Yang--Mills? Such a connection may be visible from a holographic perspective, and would be particularly interesting to determine.

In the holographic setup of AdS/CFT, it has long been known that the leading order contributions, in a large-$N$ expansion, to both $a$ and $c$ are identical \cite{Henningson:1998gx}, however it is rather rare to find holographic duals with no corrections that force the central charges to differ at a subleading order, at least, not without enhanced supersymmetry. At one-loop, corrections to the holographic central charges can come from both open string loops and closed string loops, and all these contributions must either be absent or conspire to cancel when a dual SCFT has $a = c$ exactly. Open string loops can arise from the inclusion of D-branes in the system, and these typically introduce $1/N$ corrections. Closed string loops can arise either from the $R^4$-correction to the type IIB string effective action or from the Kaluza--Klein tower on the compact space. It would be fascinating to determine if there exist holographic duals for the plethora of $\mathcal{N}=1$ SCFTs with $a = c$ that we have discovered in this paper.

\subsection{Lagrangian theories: Star/comet-shaped quivers}

To construct $\mathcal{N}=1$ SCFTs with $a=c$, we consider $(\mathcal{N}=1)$-gaugings of a collection of $\mathcal{D}_{p_i}(G)$ theories, and we discover that a necessary condition for identical central charges is that
\begin{equation}\label{eqn:gcd}
    \gcd(p_i, h_G^\vee) = 1 \,,
\end{equation}
for each $p_i$. This condition fixes that each $\mathcal{D}_{p_i}(G)$ theory lacks a Lagrangian quiver description, though we note that the converse is not true; however, it is also interesting to study the case where \emph{all} of the $\mathcal{D}_{p_i}(G)$ factors are, in fact, Lagrangian. 

In the case of the $\mathcal{N}=2$ gaugings that were used to construct the $\widehat{\Gamma}(G)$ theories, when the SCFTs are in fact Lagrangian quivers then the quivers take the form of affine ADE Dynkin diagrams, where the gauge group at each node is weighted by the associated Dynkin label. In this way, we can see that $\Gamma$ is a finite subgroup of $SU(2)$ and the quiver is the McKay graph of that finite subgroup \cite{MR604577,Kachru:1998ys}. In this paper, we have studied only those $\mathcal{N}=1$ gaugings where the resulting infrared SCFTs have $a=c$, which in the $\widehat{\Gamma}(G)$ cases involved only $\Gamma = D_4, E_6, E_7, E_8$. In this subsection, we briefly highlight the analogous Lagrangian quivers for the $\mathcal{N}=1$ gaugings discussed in this paper. 

\begin{figure}[H]
    \centering
    \begin{tikzpicture}
      \node[gaugeN1] (s0) {$pN$};
      \node[gauge] (c1) [right=1.5cm of s0] {$(p_3 - 1)\ell_3 N$};
      \node[d2] (c2) [right=0.5cm of c1] {$\cdots$};
      \node[gauge] (c3) [right=0.5cm of c2] {$2\ell_3 N$};
      \node[gauge] (c4) [right=0.5cm of c3] {$\ell_3 N$};
      \node[gauge] (t1) [above=1.5cm of c1] {$(p_1 - 1)\ell_1 N$};
      \node[d2] (t2) [right=0.5cm of t1] {$\cdots$};
      \node[gauge] (t3) [right=0.5cm of t2] {$2\ell_1 N$};
      \node[gauge] (t4) [right=0.5cm of t3] {$\ell_1 N$};
      \node[gauge] (uc1) [above=0.4cm of c1] {$(p_2 - 1)\ell_2 N$};
      \node[d2] (uc2) [right=0.5cm of uc1] {$\cdots$};
      \node[gauge] (uc3) [right=0.5cm of uc2] {$2\ell_2 N$};
      \node[gauge] (uc4) [right=0.5cm of uc3] {$\ell_2 N$};
      \node[gauge] (lc1) [below=0.4cm of c1] {$(p_4 - 1)\ell_4 N$};
      \node[d2] (lc2) [right=0.5cm of lc1] {$\cdots$};
      \node[gauge] (lc3) [right=0.5cm of lc2] {$2\ell_4 N$};
      \node[gauge] (lc4) [right=0.5cm of lc3] {$\ell_4 N$};
      \node[gauge] (b1) [below=1.5cm of c1] {$(p_5 - 1)\ell_5 N$};
      \node[d2] (b2) [right=0.5cm of b1] {$\cdots$};
      \node[gauge] (b3) [right=0.5cm of b2] {$2\ell_5 N$};
      \node[gauge] (b4) [right=0.5cm of b3] {$\ell_5 N$};
      \draw (s0.east) -- (c1.west);
      \draw (s0.60) -- (t1.west);
      \draw (s0.25) -- (uc1.west);
      \draw (s0.335) -- (lc1.west);
      \draw (s0.300) -- (b1.west);
      \draw (t1.east) -- (t2.west);
      \draw (t2.east) -- (t3.west);
      \draw (t3.east) -- (t4.west);
      \draw (b1.east) -- (b2.west);
      \draw (b2.east) -- (b3.west);
      \draw (b3.east) -- (b4.west);
      \draw (c1.east) -- (c2.west);
      \draw (c2.east) -- (c3.west);
      \draw (c3.east) -- (c4.west);
      \draw (uc1.east) -- (uc2.west);
      \draw (uc2.east) -- (uc3.west);
      \draw (uc3.east) -- (uc4.west);
      \draw (lc1.east) -- (lc2.west);
      \draw (lc2.east) -- (lc3.west);
      \draw (lc3.east) -- (lc4.west);
      \draw[dashed, ->] (s0) to[out=220, in=500, looseness=4] node[anchor=east] {$n_a$} (s0);
    \end{tikzpicture}
    \caption{The \emph{comet-shaped} quiver that depicts the Lagrangian $\mathcal{N}=1$ quiver we consider. When we write a number $M$ inside a gauge node we remind the reader that this corresponds to an $SU(M)$ gauge group. A dashed-bordered circular node denotes an $\mathcal{N}=2$ vector multiplet, a solid-bordered circular node denotes an $\mathcal{N}=1$ vector multiplet, and an undirected link between two nodes denotes an $\mathcal{N} = 2$ bifundamental hypermultiplet. The dashed self-link corresponds to $n_a$ adjoint chiral multiplets.}
    \label{fig:genericquiv}
\end{figure}
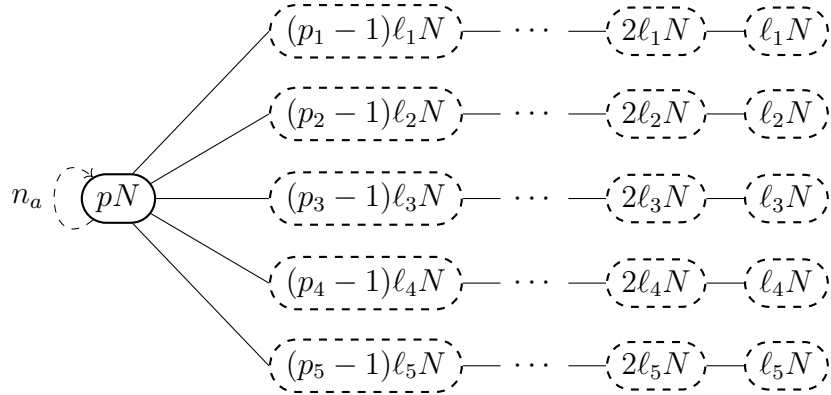

In Section \ref{sec:withchirals}, we studied the $\mathcal{N}=1$ gaugings of $\mathcal{D}_p(G)$ theories together with additional adjoint chiral multiplets on the gauge node. We make a few remarks when the $\mathcal{D}_p(G)$ theories are Lagrangian theories.
For simplicity, let us assume that $G$ is of A-type. In this case, the violation of the relationship in equation \eqref{eqn:gcd} is sufficient to ensure a Lagrangian description where we utilize
\begin{align}\label{eqn:lagexp}
\begin{aligned}
\begin{tikzpicture}
\node at (-1.7,0.3) {$\mathcal{D}_p(SU(p\ell)) =$};
\node[anchor=south west, draw, rectangle, inner sep=3pt, minimum size=5mm, text height=3mm](A0) at (0,0) {$SU(p\ell)$};
\node[anchor=south west, draw,dashed,rounded rectangle, inner sep=3pt, minimum size=5mm, text height=3mm](A1) at (2.5,0) {$SU((p-1)\ell)$};
\node(A2) at (6,0.25) {$\cdots$};
\node[anchor=south west, draw,dashed,rounded rectangle, inner sep=3pt, minimum size=5mm, text height=3mm](A3) at (7.4,-0.07) {$SU(\ell)$};
\draw (A0)--(A1)--(A2)--(A3);
\end{tikzpicture} \,\quad,
\end{aligned}
\end{align}
where a node with a dashed border indicates an $\mathcal{N}=2$ gauge node:
\begin{align}
    \begin{aligned}
    \begin{tikzpicture}
      \node[gauge] (s0) {$M$};
    \end{tikzpicture} 
    \quad\ &\raisebox{2mm}{=} \quad 
    \begin{tikzpicture}
      \node[gaugeN1] (s0) {$M$};
      \draw[dashed, ->] (s0) to[out=130, in=410, looseness=4] (s0);
    \end{tikzpicture}
    \end{aligned} \,.
\end{align}
We then consider the most generic asymptotically-free gauging setting, which involves up to five $\mathcal{D}_{p_i}(G)$ theories, and define
\begin{equation}
    p = \text{lcm}(p_1, \cdots, p_5) \qquad \text{and} \qquad \ell_i = p/p_i \,.
\end{equation}
If $G = SU(pN)$ for any positive integer $N$, then all quivers formed by gauging together such $\mathcal{D}_{p_i}(G)$ theories, possibly with additional adjoint chiral multiplets, are Lagrangian quivers. These theories will typically have $a \neq c$. We depict such quivers in Figure \ref{fig:genericquiv}. 

As shown in Figure \ref{fig:genericquiv}, these quivers are \emph{comet-shaped}, where the Lagrangian $\mathcal{D}_p(G)$ theories form the tails of the comet and the $n_a$ adjoint chiral multiplets compose the head. 
The star-shaped/comet-shaped quivers appear as the 3d ($\CN=4$) mirror theory \cite{Benini:2010uu} to a 4d $\CN=2$ theory of class $\mathcal{S}$ \cite{Gaiotto:2009hg, Gaiotto:2009we}.\footnote{See \cite{Bonelli:2019lal,Rayan:2020gzc} for a small sample of papers discussing comet-shaped quivers.} More precisely, a star/comet-shaped quiver with $n$ tails and $g$ loops corresponds to the 3d mirror theory for the class $\mathcal{S}$ theory obtained from 6d $\CN=(2, 0)$ theory compactified on a Riemann surface with genus $g$ and $n$ punctures of specific types. In our setup, the $\mathcal{D}_{p_i}(SU(pN))$ theory corresponds to the quiver tail labeled by the partition $[\ell_1 N, \ell_2 N, \ell_3 N, \cdots]$. The Higgs branch of this quiver (as a 3d $\CN=4$ theory) gives the Hitchin moduli space \cite{Gaiotto:2009hg, Benini:2010uu, Xie:2012hs}, whereas the Coulomb branch gives the Higgs branch of the 4d theory. Although what we have in Figure \ref{fig:genericquiv} looks similar to the star-shaped/comet-shaped quiver in 3d mirror theory, it is rather different. In particular, the gauge nodes in the quiver tail of a star-shaped quiver are given by $U(n)$ gauge groups, whereas in 4d we can only have $SU(n)$ type. Moreover, in 3d, the central gauge (core) node has to be gauged with an $\CN=4$ (8 supercharges) vector multiplet with possible adjoint hypermultiplets. In contrast in 4d, we consider an $\CN=1$ (4 supercharges) vector multiplet with $n_a$ adjoint chiral multiplets at the core node. 

As a generalization of the $\CN=2$ $\mathcal{D}_p(SU(p N))$ quiver tail theory, we can consider an $\CN=1$ quiver tail instead. Since the quiver tail can be obtained via nilpotent Higgsing of a linear quiver gauge theory, one can consider similar deformations to obtain $\CN=1$ quiver tails \cite{Agarwal:2014rua}. The $\CN=1$ version of the quiver tail is substantially different from the $\CN=2$ counterpart, allowing much wider set of SCFTs. It would be interesting to find an $\CN=1$ analog of $\mathcal{D}_p(G)$ Argyres--Douglas type theories that realize $a=c$ upon gauging.

\subsection{Towards a geometric construction}

Given that we construct a vast collection of 4d minimally-supersymmetric SCFTs with $a = c$, it may be possible for these theories to have both a holographic description in terms of AdS$_5$ and a top-down string theory construction. Four-dimensional $\mathcal{N}=1$ field theories can be obtained by compactifying F-theory on a non-compact elliptically-fibered Calabi--Yau fourfold, however, due to the presence of fluxes and instantons, it is not generally known when these 4d theories are conformal. Nonetheless, it is natural to ask, if we suppose that there exists an F-theory construction, whether the property of $a = c$ is encoded in the compactification data via geometry in some natural way. In this subsection, we explore some of the ways in which the F-theory geometry must be constrained if it corresponds to one of the $a = c$ theories that we consider.

Superconformal theories with $a = c$ that are formed by gauging together $N$ different $\mathcal{D}_{p_i}(G)$ theories typically have a $U(1)^{N-1}$ flavor symmetry, which is formed by the anomaly-free combinations of the $U(1)$ flavor symmetries inside the $\mathcal{N}=2$ R-symmetry from each of the $\mathcal{D}_{p_i}(G)$ theories. In F-theory compactifications, Abelian symmetries, whether gauge or flavor, arise when the elliptic fibration has a non-trivial Mordell--Weil group. The Mordell--Weil group is generated by the rational sections of the fibration. F-theory compactifications with non-trivial Mordell--Weil group have been explored in many works, for which we give a small sample here \cite{Morrison:2012ei,Borchmann:2013jwa,Cvetic:2013nia,Esole:2014dea,Lawrie:2015hia,Morrison:2014era}. The form of the geometry is then heavily constrained by the presence of these $U(1)$ symmetries, as the fibration is thus forced to have many independent rational sections. Furthermore, it is important to note that the absence of non-Abelian gauge or flavor symmetry imposes that the elliptic fibration has no reducible singular fibers supported on divisors in the base. 

The subtlety with a full Calabi--Yau fourfold compactification of F-theory is that the compactification can contain non-geometric structures, for example, arising from D3-branes wrapping compact complex surfaces inside of the base of the elliptic fibration.\footnote{Geometric and topological data relevant for elliptically-fibered Calabi--Yau compactifications of F-theory are systematically studied, for example, in \cite{Esole:2017kyr,Esole:2018bmf,Esole:2018tuz,Katz:2011qp,Gray:2013mja,Grimm:2009ef,Taylor:2015xtz}.} Due to these non-geometric effects, it is particularly difficult to find the precise condition to ensure conformality in the resulting 4d theory via non-compact fourfold compactifications which has to be imposed. To highlight the difficulty, we note that such conditions have been determined for 6d SCFTs due to the particular rigidity of threefold compactifications. Some work has been done in \cite{Apruzzi:2018oge} to explore whether certain elliptically-fibered Calabi--Yau fourfolds lead to 4d $\mathcal{N}=1$ SCFTs. 

We expect that the geometric construction would shed light upon the putative holographic understanding, or vice versa. In particular, if there exists an AdS$_5 \times X_5$ solution of Type IIB supergravity which is holographically dual to some of our gauged theories, then the $X_5$ is highly constrained by the $a = c$ condition: all string-loop contributions must cancel. An exploration of these cancellations will appear in \cite{AdS5IIB}.
Similarly, we expect that $X_5$ is related to the elliptically-fibered Calabi--Yau fourfold through which the same $\mathcal{N}=1$ theory is engineered in F-theory; in this way the $a=c$ field-theoretic condition on the $X_5$ should lead to geometric restrictions on the F-theory compactification manifold.

Instead of considering compactifications of F-theory on Calabi--Yau fourfolds, with all of the attendant complexities, we can start from the 6d $(1,0)$ SCFTs and consider a further compactification on a Riemann surface. A vast collection of six-dimensional SCFTs can be engineered in F-theory via elliptically-fibered Calabi--Yau threefolds \cite{Heckman:2013pva,Heckman:2015bfa}.\footnote{Via elliptic fibrations, the geometric engineering of F-theory provides different geometric perspectives for 6d $(1,0)$ theories (see e.g.~\cite{Esole:2015xfa,Esole:2014bka,Esole:2017rgz,Esole:2017qeh,Esole:2018mqb,Esole:2019asj,Esole:2014hya,Esole:2019hgr,Esole:2017hlw,Esole:2018csl} for some explicit geometric constructions) and we get a superconformal field theory when the Calabi--Yau threefold is non-compact and satisfies particular additional conditions.} Four-dimensional $\mathcal{N}=1$ theories can be constructed starting from these 6d $(1,0)$ theories and compactifying either on a genus $g > 1$ punctured Riemann surface, or else on a torus with flux inside of the 6d flavor symmetry. Some related aspects of the 4d theories obtained in this way have been explored in, for example, \cite{Gaiotto:2015usa,Heckman:2016xdl,Morrison:2016nrt,Razamat:2016dpl} and \cite{Bah:2017gph,Kim:2018lfo}, respectively. In such cases, we understand all the UV Abelian symmetries that may mix with the superconformal flavor symmetry under $a$-maximization, and determining the superconformal R-symmetry is straightforward. The mixing coefficients, however, are not simply related to geometric quantities and thus it obscures the origin of $a = c$ in the geometry. 

Alternatively, 4d $\mathcal{N}=1$ theories can arise from compactification of the 6d $(2,0)$ SCFTs \cite{Bah:2012dg, Beem:2012yn, Gadde:2013fma, Xie:2013gma,Bonelli:2013pva, Xie:2013rsa, Agarwal:2015vla}. Before approaching the $\mathcal{N}=1$ theories, we can consider the $\mathcal{N}=2$ theories that were obtained from gaugings of $\mathcal{D}_p(G)$ and conformal matter, and called $\widehat{\Gamma}(G)$, in \cite{Kang:2021lic}. Each of the $\mathcal{D}_p(G)$ and conformal matter factors considered in the gauging individually has a description in terms of class $\mathcal{S}$ \cite{Gaiotto:2009we,Gaiotto:2009hg}. The former are associated to spheres with one regular and one irregular puncture \cite{Xie:2012hs}, and the latter to spheres with two full punctures and one simple puncture \cite{Ohmori:2015pua,DelZotto:2015rca}. 
However, except for $\widehat{\Gamma} = \widehat{A}$, the gluing always involves the diagonal gauging of the flavor symmetries associated to at least three punctures and this is not an operation which is geometric. Indeed, for most of the $\widehat{D}_n(G)$ and $\widehat{E}_n(G)$ theories, whether Lagrangian or not, there is no known class $\mathcal{S}$ construction. 

There exists alternative geometric construction for a large class of 4d SCFTs from Calabi--Yau 3-fold singularities \cite{Xie:2015rpa, Xie:2019qmw}, some of which includes a subset of $\widehat{\Gamma}(G)$ theories \cite{Closset:2020afy}. 
In this way, the construction of $\mathcal{N}=1$ SCFTs either directly from F-theory on Calabi--Yau fourfolds, or indirectly from compactifications of the 6d $(1,0)$ SCFTs, provides a promising avenue for a top-down realization of the $a = c$ theories studied in this paper.

\section*{Acknowledgements}

We thank Richard Derryberry for a helpful discussion. M.J.K.~is partially supported by a Sherman Fairchild Postdoctoral Fellowship, the U.S.~Department of Energy, Office of Science, Office of High Energy Physics, under Award Number DE-SC0011632, and the National Research Foundation of Korea (NRF) grant NRF-2020R1A4A3079707. M.J.K., K.H.L., and J.S.~are all partly supported by the NRF grant NRF-2020R1C1C1007591. C.L.~acknowledges support from DESY (Hamburg, Germany), a member of the Helmholtz Association HGF. K.H.L.~and J.S.~are partly supported by the Breaking the Scientific Wall (BTS-W) program provided by Korea Advanced Institute of Science and Technology (KAIST). J.S.~is also supported by the Start-up Research Grant for new faculty provided by KAIST.

\bibliography{references}{}
\bibliographystyle{sortedbutpretty}

\end{document}